\documentclass[fleqn,usenatbib]{mnras}

\usepackage{newtxtext,newtxmath}

\usepackage[T1]{fontenc}
\usepackage{ae,aecompl}

\usepackage{graphicx}	% Including figure files
\usepackage{amsmath}	% Advanced maths commands
\usepackage{amssymb}	% Extra maths symbols
\usepackage[usenames,dvipsnames]{color} % Allows for comments
\newcommand{\mathcolorbox}[2]{\colorbox{#1}{$\displaystyle #2$}}

\newcommand{\brvs}{Br\"unt-V\"ais\"al\"a}
\newcommand{\alf}{Alfv\'{e}n}
\title[Convective Differential Rotation I]{Convective Differential Rotation in Stars and Planets I: Theory}

\author[Adam S. Jermyn]{
Adam S. Jermyn$^{1}$\thanks{E-mail: adamjermyn@gmail.com}
Shashikumar M. Chitre,$^{2,3}$
Pierre Lesaffre$^{4}$
and
Christopher A. Tout$^{2}$
\\
$^{1}$Center for Computational Astrophysics, Flatiron Institute, New York, New York, 10010, USA\\
$^{2}$Institute of Astronomy, University of Cambridge, Madingley Rd, Cambridge CB3 0HA, UK\\
$^{3}$Centre for Basic Sciences, University of Mumbai, India\\
$^{4}$\'{E}cole Normale Sup\'{e}rieure 24 rue Lhomond, 75231 Paris, France
}

\date{Accepted XXX. Received YYY; in original form ZZZ}

\pubyear{2020}

\begin{document}
\label{firstpage}
\pagerange{\pageref{firstpage}--\pageref{lastpage}}
\maketitle

\begin{abstract}
We derive the scaling of differential rotation in both slowly- and rapidly-rotating convection zones using order of magnitude methods.
Our calculations apply across stars and fluid planets and all rotation rates, as well as to both magnetized and purely hydrodynamic systems.
We find shear $|R\nabla\Omega|$ of order the angular frequency $\Omega$ for slowly-rotating systems with $\Omega \ll |N|$, where $N$ is the \brvs\ frequency, and find that it declines as a power-law in $\Omega$ for rapidly-rotating systems with $\Omega \gg |N|$.
We further calculate the meridional circulation rate and baroclinicity and examine the magnetic field strength in the rapidly rotating limit.
Our results are in general agreement with simulations and observations and we perform a detailed comparison with those in a companion paper.
\end{abstract}


%%%%%%%%%%% jump to intro
%%%---------- open: intro
\section{Introduction}

Differential rotation is one of the key complications in the study of stars.
It involves breaking symmetries, structure formation and both heat and momentum transport.
Importantly the origin of magnetic fields~\citep{doi:10.1146/annurev.fluid.010908.165215}, the transport of angular momentum~\citep{2014ApJ...788...93C} and the transport of material~\citep{1992A&A...253..173C} are all critically influenced by the scale and geometry of differential rotation.
Moreover, differential rotation plays a significant role in setting the spins of stellar cores, which are of significant interest for understanding the spins of white dwarfs, neutron stars and recently black holes~\citep{2019ApJ...881L...1F}.

Over the past several decades helioseismology has permitted studies of the rotation profile of the solar convection zone~\citep{1988ESASP.286..149C}.
With the passage of time the data have become more precise and detailed, providing information on the time-variability of the rotation profile~\citep{2001ApJ...559L..67A,2003ARA&A..41..599T} as well as that of its gradients~\citep{2008ApJ...681..680A}.
Similarly, related quantities such as the meridional circulation~\citep{2015ApJ...813..114R} have now been characterized.
The overall picture that has emerged for the Sun is one in which the solar differential rotation is of order $|\nabla \Omega| \approx \Omega/R$, where $\Omega$ is the angular frequency and $R$ is the distance from the spin axis.
This differential rotation reflects velocities which are large relative to the meridional circulation yet, depending on depth, may be either large or small relative to the convective velocity.
These observations present a challenge: what phenomenon sets the scale of differential rotation in the solar convection zone?

Complementing the depth of solar observations, asteroseismic observations have begun to produce information on the rotation profiles of other convecting stars~\citep{2012Natur.481...55B}.
Limits on the rotation profiles for several red giants and Sun-like stars are now known, and remarkably tell a similar picture of shear comparable to the rotation rate~\citep{2016A&A...586A..79S,doi:10.1093/mnrasl/slw171,2018Sci...361.1231B}.
This is in line with results from studies on a wider range of red giants.
These do not attempt to localize the difference rotation but find $|\nabla \Omega|$ typically of order $\Omega/R$~\citep{2015A&A...580A..96D}.
While the precision of these data and analyses continues to improve an outstanding question is whether the bulk of the differential rotation in red giants lies in their convective envelopes or in their radiative interiors~\citep{2014ApJ...788...93C}.
This highlights the importance of determining the magnitude of differential rotation in convection zones.

Unfortunately, despite these observational successes, we still lack a fully explanatory theory of differential rotation.
Early theories suggested solid body rotation~\citep{stewartson_1966}.
As turbulence and the microscopic viscosity together serve to dissipate energy, fluid bodies without any forcing are expected to come to rigid-body rotational equilibrium.
The characteristic time-scale for this is the rotation period because this is the only such scale for rotationally-driven kinetic turbulence.
This is extremely rapid in the context of stars and planets and so should preclude differential rotation.
That this is not observed is evidence of processes which inject energy into differential rotation.
For instance, in convecting bodies, turbulence may be strongly anisotropic.
Such anisotropy drives differential rotation and, if the turbulence is powered by heating rather than by the shear itself, it can maintain such a state indefinitely~\citep{1957ApJ...126..259U,1963ApJ...137..664K}.

In addition to the expectation of solid-body rotation there was also the expectation of cylindrical rotation.
This was due to the Taylor-Proudman theorem, which states that the rotation profile of an efficiently convecting (e.g. isentropic) region ought at least to have translational symmetry along the rotation axis~\citep{1897RSPTA.189..201H}.
This arises from a balance between the Coriolis and centrifugal effects, and so is very different in origin from the solid body expectation.
Despite this clean result, observations indicate that the Sun does not obey any such constraint~\citep{1998ESASP.418..759D, 1991sia..book..519G}.
This could be due to a variety of effects which the theorem neglects, including viscosity and turbulent stresses and MHD phenomena.
Additionally, convection zones are not perfectly isentropic and this leads to the so-called thermal wind correction to the Taylor-Proudman state.
One aim in this work is to determine which effects serve to break the Taylor-Proudman state and under what circumstances.

More recently, and in part owing to dramatic improvements in observational capabilities, differential rotation has garnered substantial theoretical attention.
Some authors argue that thermal wind balance and entropy gradients dominate the solar rotation profile~\citep{2006ApJ...641..618M,2012MNRAS.426.1546B}, while others have called this into question~\citep{2010A&A...510A..33B,2002ApJ...570..865B}.
Observations suggest that the thermal wind term is substantial, though there remain uncertainties as to its precise contribution~\citep{1976SoPh...46...29C,2008ApJ...673.1209R,Teplitskaya2015}.
Other models suggest that turbulent anisotropy is the most relevant factor in setting the differential rotation~\citep{1989drsc.book.....R,1993A&A...279L...1K,2013IAUS..294..399K}, and more complex models with various parameterizations have also been proposed~\citep{1989A&A...217..217T,0004-637X-808-1-35,2009SSRv..144..151B}.
One of our goals in this work is to understand which of the proposed effects matter the most and under what circumstances.

Numerical investigations of these issues have proven more successful at reproducing details of the solar rotation profile~\citep{2003ARA&A..41..599T,doi:10.1146/annurev.fluid.010908.165215}, though typically not for configurations that reflect realistic convective velocities or luminosities.
These differences may reflect uncertainties in sub-grid physics, or could be due to the fact that physically realistic resolution and diffusivites remain out of reach~\citep{Miesch2005,ASNA:ASNA201012345}.
Nevertheless, the results which have been found in this way are intriguing.
For instance, red giants are found in both simulation and asteroseismic inference to exhibit significant differential rotation, including cases where the angular velocity changes sign~\citep{2009ApJ...702.1078B}.

In this work we aim to understand the magnitude of differential rotation in the convection zones of stars and gaseous planets, and to understand how these solutions connect to the Keplerian limit of an accretion disk.
That is, we aim to determine the approximate magnitude and scaling of $|\nabla \Omega|$ in these systems.

It is important to emphasise that our arguments are purely from an order-of-magnitude scaling perspective.
In particular, we generally assume that dimensionless geometric factors are of order unity rather than being very large or small.
We believe that this is likely in most cases because the alternative is significant coincidence in the geometries of various fields which are determined by a variety of fundamentally dissimilar physical processes.
Thus, for instance, we are agnostic on whether baroclinic pumping or turbulent stresses play a greater role in the limit of slow rotation~\citep[c.f.][]{2006ApJ...641..618M} because we find that they exhibit identical scaling and are related by a dimensionless factor of order unity.

We further caution that there remain significant uncertainties in the precise scaling of turbulent velocities and stresses.
We have done our best to estimate these scalings from a combination of symmetry arguments and mixing length approaches.
The agreement we find with observations and simulations in the companion manuscript suggests that these tools have been useful, but they are inherently simplifications and we feel compelled to point out their limitations.

We begin in Section~\ref{sec:assump} with a discussion of our assumptions.
In Section~\ref{sec:vort} we examine the vorticity equation and derive the form which we use in all subsequent analysis.
We then consider in turn magnetic fields (Section~\ref{sec:mag}), the condition of thermal equilibrium (Section~\ref{sec:merid}) and the thermal wind contribution (Section~\ref{sec:thermwind}), deriving helpful expressions relating the magnitudes of different effects.
In Section~\ref{sec:stiff} we introduce an asymptotic scaling approach which helps to organise the remainder of the calculation.

The remaining Sections focus on our key results.
In Sections~\ref{sec:slow},~~\ref{sec:fast_hydro} and~\ref{sec:fast} we examine the differential rotation in both rapidly and slowly rotating convection zones in both the MHD and non-magnetic limits.
In Appendices~\ref{sec:cascade} and~\ref{sec:breakup} we show how these results relate to the inverse cascade in density-stratified systems and how the limit of rapid rotation transitions into a Keplerian state.
These are separated from the main text because their results are applicable in somewhat more limited circumstances.

Taken together our results provide a unified theory of the magnitude of convective differential rotation which covers the full range from accretion discs to gas planets to the most massive stars.
We provide a detailed comparison of our model with both observations and simulations in a companion manuscript, so we conclude by discussing the limitations of our analysis (Section~\ref{sec:limitations}), summarizing our results (Section~\ref{sec:summ}, Table~\ref{tab:summary}) and commenting on their astrophysical implications (Section~\ref{sec:conc}).%%%---------- close: intro

%%%%%%%%%%% jump to assumptions
%%%---------- open: assumptions
\section{Assumptions}
\label{sec:assump}

For simplicity, we make a few assumptions.
\begin{enumerate}
	\item Dimensionless factors arising from geometry are of order unity unless symmetries require them to be otherwise.
	\item All external perturbing forces, such as tides or external heating, are negligible in the regions of interest.
	\item The material is non-degenerate, compressible and not radiation-dominated.
	\item All microscopic (i.e. non-turbulent) diffusivities are negligible, such that
	\begin{enumerate}
	\item convection is efficient, so the gas is nearly isentropic,
	\item the Reynolds and Rayleigh numbers are much larger than critical, and
	\item magnetohydrodynamical processes are ideal.
	\end{enumerate}
	\item The system is axisymmetric in a time-averaged sense.
	\item Convection is subsonic.
	\item The system is chemically homogeneous.
\end{enumerate}
Small violations of these assumptions do not undermine our conclusions.
For instance so long as the entropy is logarithmic in pressure and density, and the sound speed is of order $\sqrt{P/\rho}$, corrections owing to radiation pressure and degeneracy are not a problem.

Similarly, the assumption of axisymmetry is meant not in each instant but rather in a time-averaged sense.
This is a much weaker condition and importantly does not run afoul of Cowling's anti-dynamo theorem~\citep{1933MNRAS..94...39C,1955ApJ...122..293P}.
So for example turbulence and dynamo cycles may produce temporary deviations from axisymmetry but the long-term average behaviour must be axisymmetric.

The most important assumption lies in our treatment of geometric factors.
We do not treat the effects of spherical geometry in detail: we approximate latitude-dependent effects with their averages over $\theta$, and take angular derivatives to produce factors of $r^{-1}$.
Moreover, aside from considering the ratio of the pressure scale height $h$ to the radius $r$ we do not consider the effects of the aspect ratio or depth of the convection zone, and do not differentiate between spherical or shellular geometries.
On the other hand we are highly concerned with the consequences of spherical symmetry and the ways in which it breaks.
For instance a non-rotating self-gravitating system with no external or fossil magnetic field is spherically symmetric and so, even though the convective stresses do not vanish, the angular momentum they transport does.
Hence in the slowly-rotating regime there is a small, rotation-dependent geometric factor associated with rotation breaking this symmetry which relates the stress that transports angular momentum to that which does not.
We pay significant attention to such terms.
More broadly, we effectively assume that, apart from any symmetries, the boundary conditions and geometry of the system are relatively generic.
A consequence of this is that we generally avoid assuming that terms in the solutions are tuned to be irrelevant or that they contain geometric factors which scale with rotation rate.%%%---------- close: assumptions

%%%%%%%%%%% jump to vorticity
%%%---------- open: vorticity

\section{Vorticity}
\label{sec:vort}

First, we derive the laws governing the evolution of angular momentum in axisymmetric systems, paying particular attention to the distinction between meridional and longitudinal components.
We then argue that these systems are likely to be in a quasi-steady state.

\begin{figure}
\includegraphics[width=0.47\textwidth]{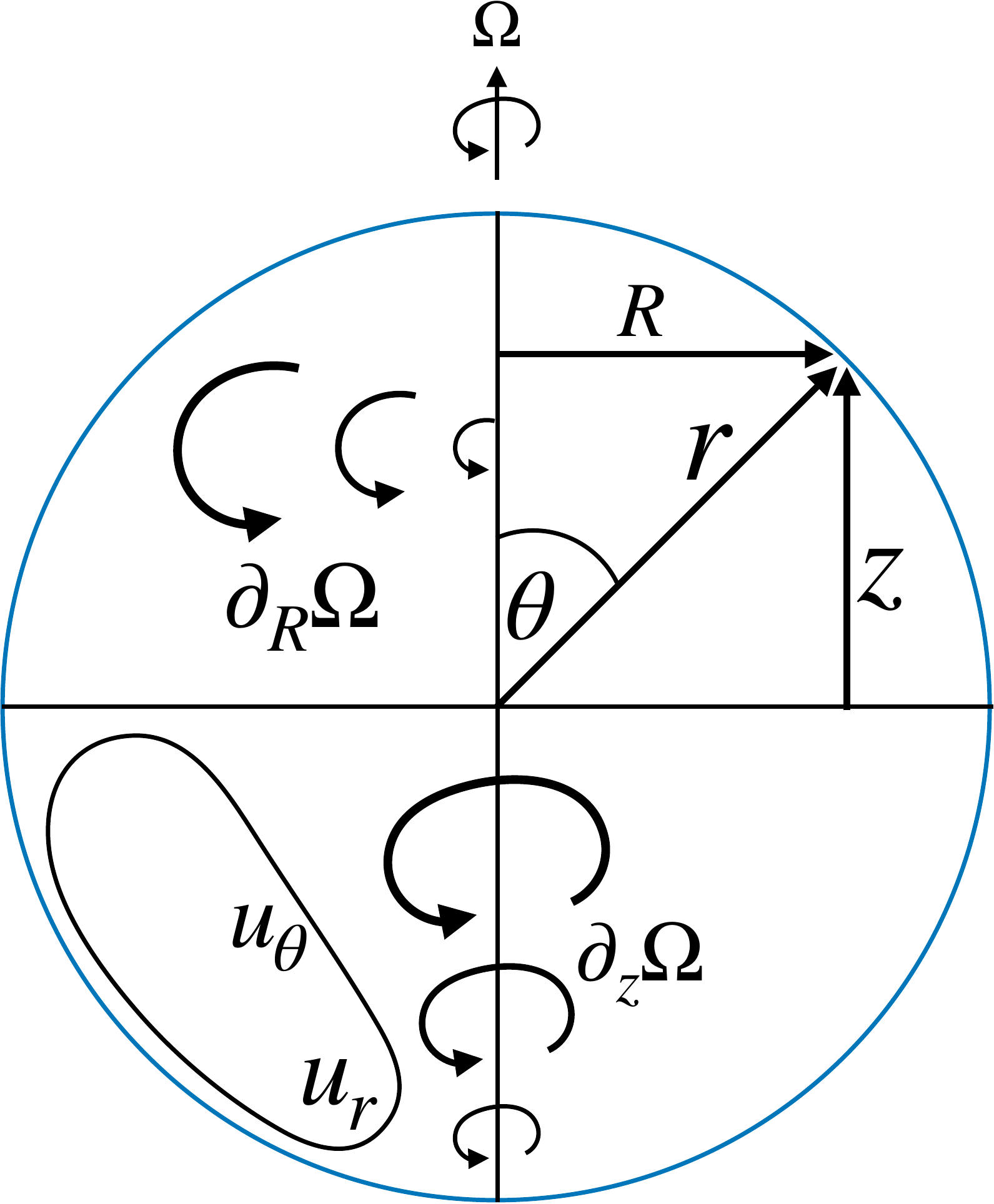}
\caption{The rotation, coordinate system and differential rotation are shown schematically: (top) the mean angular velocity $\Omega$; (upper-right) the cylindrical radius $R$, vertical direction along the rotation axis $z$, spherical radius $r$ and polar angle $\theta$; (upper-left) an example of cylindrical radial differential rotation ($\partial_R \Omega$); (lower) an example of cylindrical vertical differential rotation ($\partial_z\Omega$); (lower-left) an example of a meridional circulation $\boldsymbol{u}$.}
\label{fig:schema_coords}
\end{figure}

We begin by defining vorticity of the fluid as
\begin{align}
	\boldsymbol{\omega} \equiv \nabla \times \boldsymbol{\varv},
	\label{eq:vort}
\end{align}
where $\boldsymbol{\varv}$ is the velocity.
The vorticity is the angular velocity of the fluid about a point, and so is closely related to the differential rotation.
In particular in the limit where rotation dominates $\boldsymbol{\varv} = R \Omega \boldsymbol{e}_\phi$ and
\begin{equation}
	\boldsymbol{\omega} = \Omega \boldsymbol{e}_z + R\nabla\times(\Omega \boldsymbol{e}_\phi),
\end{equation}
where $\boldsymbol{\Omega}$ is the local angular velocity about the $z$ axis and $R$ is the cylindrical radial coordinate (Fig.~\ref{fig:schema_coords}).
We do not impose this limit but it is worth keeping in mind because it gives an intuitive connection between vorticity and the force of rotation.

By taking the curl of the Navier-Stokes equation we have that, in the absence of external forcing (e.g. tides), the vorticity evolves according to
\begin{equation}
\frac{\partial \boldsymbol{\omega}}{\partial t} = \boldsymbol{\omega}\cdot\nabla\boldsymbol{\varv} - \boldsymbol{\omega}\nabla\cdot\boldsymbol{\varv} - \boldsymbol{\varv}\cdot\nabla\boldsymbol{\omega} + \frac{1}{\rho^2}\nabla\rho\times\nabla P + \nabla\times\left(\frac{1}{\rho}\nabla\cdot\mathbfss{T}\right)	+\nabla\times\left(\frac{\boldsymbol{F}_{\! B}}{\rho}\right),
\label{eq:vorticity}
\end{equation}
where $\boldsymbol{F}_{\! B}$ is the magnetic force, $\rho$ is the density, $p$ is the pressure and $\mathbfss{T}$ is the turbulent fluid stress such that $\mathsf{T}_{ij}$ is the flux of $\boldsymbol{e}_i$ momentum along the direction $\boldsymbol{e}_j$, such that
\begin{align}
	\left(\nabla\cdot\mathbfss{T}\right)_i = \sum_{j} \frac{\partial \mathsf{T}_{ij}}{\partial x_j}.
\end{align}

We are interested in axisymmetric systems, where there is a natural distinction between the meridional and longitudinal components of the flow  (Fig.~\ref{fig:schema_coords}).
To make this explicit we write the meridional flow as
\begin{equation}
	\boldsymbol{u}(R,z) \equiv \boldsymbol{\varv}(R,z) - \Omega(R,z) R \boldsymbol{e}_\phi.
	\label{eq:meridional_flow_breakout}
\end{equation}
Inserting this into equation~\eqref{eq:vorticity}, we then obtain (Appendix~\ref{appen:vort})
\begin{align}
\frac{\partial \boldsymbol{\omega}}{\partial t} = &\,\,\boldsymbol{\omega}\cdot\nabla\boldsymbol{u} - \boldsymbol{\omega}\nabla\cdot\boldsymbol{u} - \boldsymbol{u}\cdot\nabla\boldsymbol{\omega}+\boldsymbol{e}_\phi (\boldsymbol{\omega}\cdot\nabla)(\Omega R)-\boldsymbol{e}_\phi\Omega \omega_R\nonumber\\
& + \frac{1}{\rho^2}\nabla\rho\times\nabla P + \nabla\times\left(\frac{1}{\rho}\nabla\cdot\mathbfss{T}\right)+\nabla\times\left(\frac{\boldsymbol{F}_{\! B}}{\rho}\right).
\label{eq:vorticity1}
\end{align}
The first line of this equation describes kinematic effects associated with the rotation and circulation while the second describes the effects of the thermal wind, turbulent stresses and magnetic stresses.

Astrophysical systems are likely to be near angular momentum equilibrium because the time-scale over which shear turbulence transports momentum is quite short.
To see this note that the diffusivity of shear turbulence over the whole system is of order\footnote{This estimate is for neutrally or unstably stratified media and does not generalize to the case of stable stratification.}
\begin{align}
	\nu \approx R^2 |\nabla \boldsymbol{\varv}|,
\end{align}
where $R$ is the radius of the star.
This is because the shear time-scale is set by $|\nabla \boldsymbol{\varv}|$ and its associated length-scale is set by the distance over which it persists.
The diffusive timescale associated with a shear is therefore
\begin{align}
	\tau_{\mathrm{diff}} \approx \frac{R^2}{\nu} \approx |\nabla\boldsymbol{\varv}|^{-1},
	\label{eq:tDiff}
\end{align}
which means that
\begin{align}
	\frac{d|\nabla \boldsymbol{\varv}|}{dt} \approx -\frac{|\nabla \boldsymbol{\varv}|}{\tau_{\mathrm{diff}}} \approx |\nabla \boldsymbol{\varv}|^2,
\end{align}
so at times longer than the initial $\tau_{\mathrm{diff}}$
\begin{align}
|\nabla \boldsymbol{\varv}| \propto (|\nabla \boldsymbol{\varv}|t)^{-1}.
\end{align}
What this implies is that transients ring down on a time-scale comparable to their size.
So after astrophysical time-scales, transient shears are likely very small, and the system is well-approximated by instantaneous momentum equilibrium.
This applies even in the presence of secular evolution owing to nuclear processes and wind losses and suggests that those effects are generally not enough to violate angular momentum equilibrium.

There are two cases in which this argument fails.
The first is in accretion discs, where the diffusivity is suppressed relative to equation~\eqref{eq:tDiff}, resulting in less relative angular momentum transport and hence longer equilibration time-scales.
However, in this case transients must ring--down on time-scales of order $\Omega^{-1}$ because non-Keplerian motion results in the fluid centrifugally adjusting its orbit on this time-scale.
So, once more we find that transient effects in the differential rotation decay quickly.

The second case is in systems involving waves and turbulence.
Equation~\eqref{eq:vorticity1} typically supports linear oscillatory motions, such as \alf\ waves and sound waves, as well as instabilities such as those associated with a dynamo~\citep{ASNA:ASNA201012345} and convection~\citep{1958ZA.....46..108B}.
We take these motions to have been averaged over time, such that mean effects appear in the stress tensor $\mathbfss{T}$ and so we neglect them wherever they appear explicitly in the vorticity equation.
We also specifically neglect gravity waves because these are evanescent in convection zones~\citep{0004-637X-796-1-17}.

We now set the time derivative in equation~\eqref{eq:vorticity1} to zero and find
\begin{align}
0 = &\,\,\boldsymbol{\omega}\cdot\nabla\boldsymbol{u} - \boldsymbol{\omega}\nabla\cdot\boldsymbol{u} - \boldsymbol{u}\cdot\nabla\boldsymbol{\omega}+\boldsymbol{e}_\phi \boldsymbol{\omega}\cdot\nabla(\Omega R)-\boldsymbol{e}_\phi\Omega \omega_R\nonumber\\
& + \frac{1}{\rho^2}\nabla\rho\times\nabla P + \nabla\times\left(\frac{1}{\rho}\nabla\cdot\mathbfss{T}\right)	+ \nabla\times\left(\frac{\boldsymbol{F}_{\! B}}{\rho}\right).
\label{eq:vorticity3}
\end{align}
Furthermore, in steady state conservation of mass requires
\begin{align}
	\nabla\cdot(\rho \boldsymbol{u}) = 0
	\label{eq:masscons}
\end{align}
Inserting this into the second term of equation~\eqref{eq:vorticity3} and separating this vector-valued equation into its meridional and $\boldsymbol{e}_\phi$ components we find
\begin{align}
	\label{eq:vortm0}
	0 =\ &\boldsymbol{\omega}_m\cdot\nabla \boldsymbol{u} + \boldsymbol{\omega}_m \boldsymbol{u}\cdot\nabla\ln\rho + \boldsymbol{u}\cdot\nabla\boldsymbol{\omega}_m+\left[\nabla\times\left(\frac{1}{\rho}\nabla\cdot \mathbfss{T}\right)\right]_m\nonumber\\
	& + \left[\nabla\times\left(\frac{\boldsymbol{F}_{\! B}}{\rho}\right)\right]_m\\
\intertext{and}
	\label{eq:vortp0}
	0 =\ &R^{-1} \omega_\phi u_R + \omega_\phi \boldsymbol{u}\cdot\nabla\ln\rho + \boldsymbol{u}\cdot\nabla\boldsymbol{\omega}_\phi + \boldsymbol{\omega}\cdot\nabla(\Omega R) - \Omega \boldsymbol{\omega}_R\nonumber\\
	& + \rho^{-2}\boldsymbol{e}_\phi\cdot\nabla P \times \nabla \rho + \boldsymbol{e}_\phi\cdot\nabla\times\left(\frac{1}{\rho}\nabla\cdot \mathbfss{T}\right) + \boldsymbol{e}_\phi\cdot\nabla\times\left(\frac{\boldsymbol{F}_{\! B}}{\rho}\right).
\end{align}
where we denote the projection of a vector into the meridional plane by the subscript $m$, i.e.
\begin{equation}
	\boldsymbol{u} \equiv \boldsymbol{\varv}_m.
\end{equation}
%%%---------- close: vorticity

%%%%%%%%%%% jump to magnetic
%%%---------- open: magnetic

\section{Magnetic Fields}
\label{sec:mag}

The appearance of the magnetic force in equation~\eqref{eq:vorticity3} means that to close this equation we must address the source of magnetism.
There are typically three sources of magnetic fields in astrophysical contexts, fossil fields, turbulent dynamos and externally imposed fields.

A fossil field is present at the time of formation of the body.
From the perspective of a scaling analysis there is little difference between a fossil field and an externally imposed field: both may break symmetries, including axisymmetry, and in both cases there is no guarantee of a relationship between the magnitude of the field and other properties of the body.
Hence, they may be analyzed together.

Such an analysis awaits future study.
For now, we consider the many cases in which the dominant magnetic field is generated by turbulent dynamo processes.
In such cases the field obeys the same symmetry, on average, as the turbulence which drives it\footnote{The only exception to this is if the dynamo spontaneously breaks one of the symmetries of the system in a time-averaged sense. Hence it is not sufficient that the solar magnetic field at any moment points in a particular direction because when averaged over the solar cycle the field vanishes. As far as we are aware, no such spontaneous symmetry breaking has been observed in stars or seen in MHD simulations, so we assume that this does not occur.}, so its contribution vanishes in highly symmetric situations just as does the contribution of the turbulence.
What remains then is due to the fluctuations, which we absorb into the turbulent stress, such that
\begin{align}
	\label{eq:vortm1}
	0 =\ &\boldsymbol{\omega}_m\cdot\nabla \boldsymbol{u} + \boldsymbol{\omega}_m \boldsymbol{u}\cdot\nabla\ln\rho + \boldsymbol{u}\cdot\nabla\boldsymbol{\omega}_m+\left[\nabla\times\left(\frac{1}{\rho}\nabla\cdot \mathbfss{T}\right)\right]_m\\
\intertext{and}
	\label{eq:vortp1}
	0 =\ &R^{-1} \omega_\phi u_R + \omega_\phi \boldsymbol{u}\cdot\nabla\ln\rho + \boldsymbol{u}\cdot\nabla\boldsymbol{\omega}_\phi + \boldsymbol{\omega}\cdot\nabla(\Omega R) - \Omega \boldsymbol{\omega}_R\nonumber\\
	& + \rho^{-2}\boldsymbol{e}_\phi\cdot\nabla P \times \nabla \rho + \boldsymbol{e}_\phi\cdot\nabla\times\left(\frac{1}{\rho}\nabla\cdot \mathbfss{T}\right),
\end{align}
where $\mathbfss{T}$ now includes the turbulent magnetic stress
\begin{align}
	\mathbfss{T}_{B} \equiv \frac{1}{4\pi}\left(\boldsymbol{B}\otimes\boldsymbol{B}-\frac{1}{2}\mathbfss{I} B^2\right).
\end{align}
So what we must determine is how this compares with the characteristic scale of the fluid stress.
There are two regimes of interest, namely that of slow rotation and that of rapid rotation, which we discuss in sections~\ref{sec:slow} and~\ref{sec:fast}.
%%%---------- close: magnetic

%%%%%%%%%%% jump to thermal_eq
%%%---------- open: thermal_eq

\section{Thermal Equilibrium}
\label{sec:merid}

Equations~\eqref{eq:vortm1} and~\eqref{eq:vortp1} involve the meridional circulation and the baroclinicity, so we must close the system of equations with a study of heat transport.
We do so through the equation of thermal equilibrium, which reads
\begin{align}
	\mu^{-1} \rho c_p T \boldsymbol{u}\cdot\nabla s + \nabla\cdot\boldsymbol{F} = 0,
	\label{eq:heat0}
\end{align}
where $\mu$ is the mean molecular weight, $\rho$ is the density, $c_p$ is the specific heat at constant pressure, $T$ is the temperature, 
\begin{align}
	s \equiv \frac{1}{\gamma-1}\left(\ln P - \gamma \ln \rho\right)
	\label{eq:s}
\end{align}
is a dimensionless entropy for an ideal gas (see appendix~\ref{appen:entropy}) and
\begin{align}
	\boldsymbol{F} = \mu^{-1} \rho c_p T \mathbfss{Q}\cdot\nabla s
	\label{eq:QF}
\end{align}
is the convective flux with diffusivity tensor $\mathbfss{Q}$.
For an ideal gas $P \propto \mu^{-1}\rho c_p T$ so equation~\eqref{eq:heat0} may be written as
\begin{align}
	P \boldsymbol{u}\cdot\nabla s + \nabla\cdot\left(P \mathbfss{Q}\cdot\nabla s\right) = 0.
	\label{eq:heat1}
\end{align}
In the non-rotating limit this possesses barotropic solutions with $\nabla s$ and $\nabla P$ both radial.
Outside of this limit that is generally not true, both because $\nabla P$ is distorted by centrifugal forces~\citep{1929MNRAS..90...54E} and because $\mathbfss{Q}$ becomes anisotropic~\citep{2013IAUS..294..399K}.

\subsection{Coordinate System}

\begin{figure}
\includegraphics[width=0.47\textwidth]{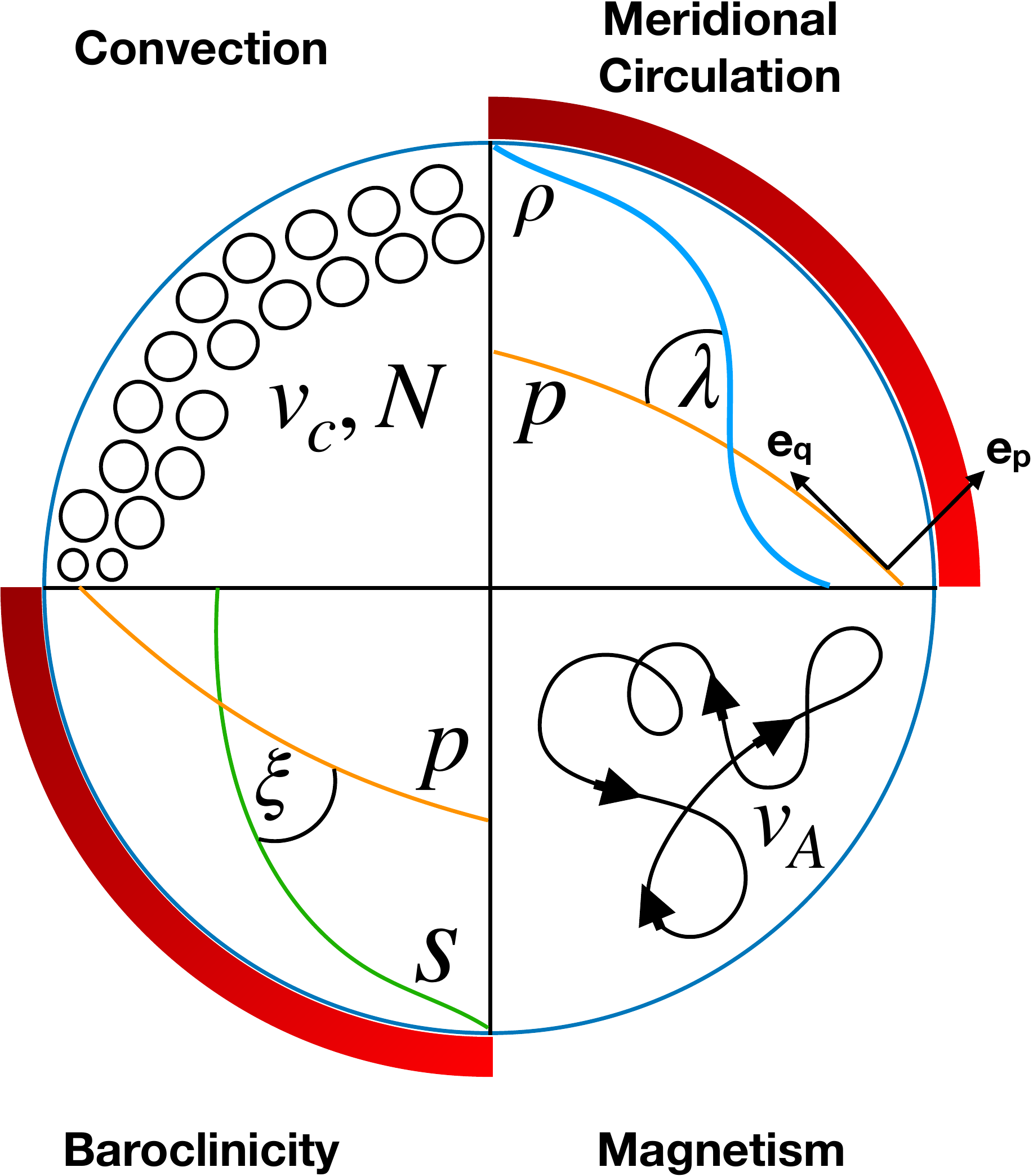}
\caption{Key concepts in our theory are shown schematically: (upper-left) turbulent eddies move at the convection speed $\varv_c$ and with the \brvs\ frequency $N$; (lower-left) surfaces of constant pressure and entropy meet at an angle of approximately the baroclinicity $\xi$, resulting in surface temperature variations; (lower-right) a magnetic field with \alf\ velocity $\varv_{\rm A}$; (upper-right) surfaces of constant pressure and density meet at an angle of approximately $\lambda$, the unit vector $\boldsymbol{e}_p$ points along the pressure gradient and the unit vector $\boldsymbol{e}_q$ is perpendicular to it.}
\label{fig:schema_properties}
\end{figure}

To proceed we define $\boldsymbol{e}_p$ to be the unit vector along the pressure gradient and $\boldsymbol{e}_q \equiv \boldsymbol{e}_\phi \times \boldsymbol{e}_p$ to be a unit vector perpendicular to $\boldsymbol{e}_p$ in the meridional plane (Fig.~\ref{fig:schema_properties}, upper-right).
We denote components of vectors by the subscripts $p$ and $q$ to mean the components along these unit vectors.
It is also useful to define the baroclinicity
\begin{align}
	\xi \equiv \frac{\boldsymbol{e}_\phi \cdot \left(\nabla \ln P \times \nabla s\right)}{|\nabla \ln P| |\nabla s|},
	\label{eq:xi}
\end{align}
which measures the extent to which the pressure and entropy gradients are misaligned (Fig.~\ref{fig:schema_properties}, lower-left).
When $\nabla P$ and $\nabla s$ are nearly aligned, $\xi$ measures the small angle between them.
When they are further misaligned it approaches $\pm 1$.
For convenience, we also define
\begin{align}
	\bar{\xi} \equiv \pm\sqrt{1-\xi^2},
\end{align}
where the branch of the square root is taken such that
\begin{align}
	\nabla s = |\nabla s| \left(\boldsymbol{e}_p\bar{\xi} + \boldsymbol{e}_q \xi\right).
	\label{eq:grad_s}
\end{align}
With this definition
\begin{align}
	\boldsymbol{u} = \boldsymbol{e}_p u_p + \boldsymbol{e}_q u_q.
\end{align}

We can relate $\xi$ to the pressure and density gradients by defining
\begin{align}
	\lambda \equiv \frac{|\nabla \ln P \times \nabla \ln \rho|}{|\nabla \ln P| |\nabla \ln \rho|},
	\label{eq:lambda}
\end{align}
which is directly proportional to the thermal wind term in equation~\eqref{eq:vortp1} (Fig.~\ref{fig:schema_properties}, upper-right).
When $\nabla P$ and $\nabla \rho$ are nearly aligned, $\lambda$ measures the small angle between them.
When they are orthogonal $\lambda$ approaches unity.
In analogy with $\bar{\xi}$ we also define
\begin{align}
	\bar{\lambda} \equiv \pm\sqrt{1-\lambda^2},
	\label{eq:bar_lambda}
\end{align}
where the sign is chosen to match that of $\bar{\xi}$.

\subsection{Meridional Circulation}

We next consider the meridional circulation.
This operates under two constraints: conservation of mass (equation~\ref{eq:masscons}) and thermal equilibrium (equation~\ref{eq:heat1}).
In Appendix~\ref{subsec:masscons} we expand the first of these in the coordinate system we have just defined and obtain (equation~\ref{eq:upq2})
\begin{align}
	|u_p| \approx \left(\lambda + \frac{h}{r}\right) |u_q|,
	\label{eq:upq}
\end{align}
where $h$ is the pressure scale height defined in equation~\eqref{eq:h} as
\begin{align}
	h &\equiv |\nabla \ln P|^{-1} = \frac{P}{\rho g}
\end{align}
and $g$ is the acceleration owing to gravity.

Equation~\eqref{eq:upq} says that the vertical flow $u_p$ is suppressed relative to the horizontal flow $u_q$ by a factor involving the density stratification.
In Appendix~\ref{subsec:thermo} we combine this with the condition of thermal equilibrium and find (equation~\ref{eq:uu})
\begin{align}
	\left(\xi + \frac{h}{r}\right) u + \frac{\left|\nabla\cdot\left(P \mathbfss{Q}\cdot\nabla s\right)\right|}{P |\nabla s|} \approx 0,
	\label{eq:u}
\end{align}

Equation~\eqref{eq:u} relates the magnitude of the meridional circulation to the conditions of thermal equilibrium, which in turn are related to the baroclinicity by the diffusivity tensor.
This does not intrinsically mean that thermal equilibrium drives the circulation, just that thermal equilibrium demands that this equation be satisfied.
Thus, for instance, if the vorticity equation is more sensitive to the meridional circulation than the momentum equation it could be that the momentum equation drives a circulation, in which case the causation runs from left-to-right and the circulation determines the baroclinicity~\citep{doi:10.1146/annurev.fluid.010908.165215,2009SSRv..144..151B}.
Likewise, in the reverse case baroclinicity drives and determines the circulation~\citep{1929MNRAS..90...54E, 1982PASJ...34..257O,1998A&A...334.1000M}.
In fact we show that both possibilities likely occur occur depending on the context.
A related result from recent simulations is that, in the slow-rotation limit, these two effects actually scale with rotation in a similar manner~\citep{doi:10.1146/annurev.fluid.010908.165215}.
The momentum imbalance generally provides a stronger impetus.
This may be why different approaches, focusing on one or the other, have made similar predictions for the circulation rate.%%%---------- close: thermal_eq

%%%%%%%%%%% jump to thermal_wind
%%%---------- open: thermal_wind

\section{Thermal Wind}
\label{sec:thermwind}

We now turn to the thermal wind balance, which enters via the baroclinic term $\rho^{-2}(\nabla P\times\nabla \rho)_\phi$ appearing in equations~\eqref{eq:vortp1}.
Noting that $\nabla P$ and $\nabla \rho$ lie in the meridional plane we find
\begin{align}
	\label{eq:tw0}
	|\rho^{-2}(\nabla P\times\nabla \rho)_\phi| &= \frac{P}{\rho}|\nabla \ln P \times \nabla \ln \rho|\\
	\label{eq:tw1}
	&=\gamma N^2\frac{\xi}{\bar{\xi}},
\end{align}
See Appendix~\ref{subsec:tw} for more detail on the algebra between equations~\eqref{eq:tw0} and~\eqref{eq:tw1}.
Here $N$ is the \brvs\ frequency defined in equation~\eqref{eq:brvs} as (Fig.~\ref{fig:schema_properties}, upper-left)
\begin{align}
	N^2 \equiv -\frac{\gamma-1}{\gamma} \boldsymbol{g}\cdot\nabla s.
\end{align}
Because $\gamma$ and $\bar{\xi}$ are of order unity we drop these and obtain
\begin{align}
	|\rho^{-2}(\nabla P\times\nabla \rho)_\phi| \approx N^2 \xi,
	\label{eq:thermwind}
\end{align}%%%---------- close: thermal_wind

%%%%%%%%%%% jump to stiffness
%%%---------- open: stiffness

\section{Asymptotic Analysis}
\label{sec:stiff}

Our aim is to estimate the asymptotic behavior of the differential rotation as $\Omega/|N| \rightarrow 0$ and as $\Omega/|N| \rightarrow \infty$.
By approximating gradients with appropriate length-scales we obtain equations of the form
\begin{align}
	\sum_{k} x_{i,k} \prod_{l} \phi_{l}^{\alpha_{i,k,l}} = 0,
	\label{eq:schematic}
\end{align}
where $x_{i,k}$ are coefficients of order unity and $\phi_{l}$ are the variables which parameterize the problem.
In particular $\{\phi_{l}\}$ includes $\Omega$, $\xi$, $u$, $R\partial_R \Omega$ and $R\partial_z \Omega$, each non-dimensionalized by appropriate factors of $h$ and $|N|$.
The equations are derived from thermal equilibrium~\eqref{eq:heat0}, meridional vorticity balance~\eqref{eq:vortm1} and azimuthal vorticity balance~\eqref{eq:vortp1}, and reflect their scaling in the relevant limits.
The exponents $\alpha_{i,k,l}$ are independent of the rotation rate.
We solve these equations for $\xi$, $u$, $R\partial_R \Omega$ and $R\partial_z \Omega$ as functions of $\Omega$.
Note that because the meridional component of the vorticity equation has two components we have four equations and four unknowns, so the system is determined.

The solution $\{\phi_l(\Omega)\}$ may be expanded as a Puiseux series~\citep{2008arXiv0811.0414A}.
In the limit of asymptotically large or small $\Omega$ these series are well approximated by power-laws.
That is,
\begin{align}
	\phi_{l} \propto \left(\frac{\Omega}{|N|}\right)^{\beta_{l}}.
\end{align}
Our aim is to determine the exponents $\beta_l$.
To do this we note that asymptotic solutions to equation~\eqref{eq:schematic} must have at least two of the terms in the equation be of comparable magnitude.
Were this not the case there would be a single term which is asymptotically larger than all the others, which would preclude the sum of all terms vanishing.
We therefore seek the $\beta_l$ which cause each equation to have two or more terms which are larger than the rest and scale in the same manner as one another.

It is important to note that there need not be a single unique solution.
This may arise either because there is an unphysical branch or because either of multiple terms suffices to balance another.
In the former case we employ physical arguments to eliminate the extraneous solutions.
In the latter case we assume that solutions are not fine-tuned to select just one of the various terms which may balance, so that all such terms exhibit the same asymptotic behavior.
So for instance in the slowly-rotating limit we shall find that either baroclinicity or a meridional circulation suffices to balance the equation of thermal equilibrium, and that these terms enter in the same manner in the vorticity equation.
Without a reason to believe that the system fine-tunes to have the baroclinicity vanish and the meridional circulation carry the full burden, or vice-versa, we assume that they share the responsibility.
Note that this means we often cannot say anything about the relative signs of the $\{\phi_l(\Omega)\}$ we obtain, because there could be more than two terms involved in the overall heat and momentum balance.

In each limit we compute the terms and present an heuristic argument for the scalings $\beta_l$.
To confirm our results we also perform a search over all possible pairs of terms in each equation which might balance and compute the required asymptotic scalings\footnote{The software used to perform this search is available at \url{https://doi.org/10.5281/zenodo.3967763}.}.

We discard solutions for which the chosen pair is not dominant.
Because each $\phi$ vanishes as $\Omega\rightarrow 0$ by symmetry in the slowly rotating limit we discard any solution with any $\beta_l\leq 0$.
Similarly, in the rapidly-rotating limit we discard any solution for which $u/h|N|$ or $|R\nabla\Omega|/|N|$ diverge as $\Omega \rightarrow \infty$.
This is because shears are dissipative, with energy loss per unit mass
\begin{align}
	\dot{E} \approx h^2 \omega^{3},
\end{align}
where $\omega$ is the shear time-scale.
The only source of energy in these systems is the entropy gradient, which may act either through baroclinic pumping or convective forcing.
This has characteristic power scale
\begin{align}
	\dot{E} \approx h^2 |N|^3,
\end{align}
so any shear with $\omega > |N|$ dissipates faster than it is forced and cannot be sustained over long time-scales.%%%---------- close: stiffness

%%%%%%%%%%% jump to slow_rot
%%%---------- open: slow_rot
\section{Slow Rotation}
\label{sec:slow}

In the special case of a non-rotating body with no fossil field, every term in the vorticity equation vanishes.
This follows from symmetry because the only preferred direction is radial and this implies that the system is spherically symmetric.
All vector fields of interest must therefore be radial\footnote{The only exception to this occurs if there is turbulence which exhibits spontaneous symmetry breaking, as mentioned previously.}.
Every term in the vorticity equation results from the curl of a vector field and the curl of a radial field with spherical symmetry vanishes.
Consequently every term in equation~\eqref{eq:vorticity3} vanishes.
This provides a useful starting point for perturbation theory in the slow-rotation limit.
Note that here slow is with respect to the \brvs\ frequency $|N|$, so we assume that $\Omega \ll |N|$.

\subsection{Magnetic Field ($\Omega \ll |N|$)}

When the rotation is slow relative to the convective turnover time, the field approaches equipartition with the turbulent flow~\citep{2000A&A...360..335A,2000RvMP...72.1081R,2006GeoJI.164..467S} and the Alfv\'{e}n speed (Fig.~\ref{fig:schema_properties}, lower-right)
\begin{align}
	\varv_\mathrm{A} \equiv \frac{B}{\sqrt{4\pi \rho}}
\end{align}
is comparable to the convection speed.
This is what is typically found in simulations~\citep{0004-637X-803-1-42,2011IAUS..271..361A} even up to rotation rates comparable to the turnover time~\citep{0004-637X-777-2-153}.
So in this regime the magnetic contribution to the turbulent stress scales with the kinetic term and we can focus on the latter.

\subsection{Stress ($\Omega \ll |N|$)}
\label{sec:slowstress}

We turn now to the contributions from the convective stress.
When the rotation is slow the turbulence is primarily convective, with characteristic length scale $h$ and characteristic time scale $|N|^{-1}$.
As a result~\citep{1958ZA.....46..108B}
\begin{align}
	\mathsf{T} \approx \rho h^2 |N|^2,
	\label{eq:Tslow}
\end{align}
where $\mathsf{T}$ is the typical magnitude of entries in $\mathbfss{T}$.

In the non-rotating limit the stress is constrained by spherical symmetry to be of the form
\begin{align}
	\mathbfss{T} &\approx 
	\begin{pmatrix}
\mathsf{T}_{rr} & 0 & 0 \\ 
0 & \mathsf{T}_{\theta\theta} & 0 \\
0 & 0 & \mathsf{T}_{\phi\phi} 
\end{pmatrix},
\end{align}
where we have written this tensor in spherical coordinates and the diagonal terms are of the same order of magnitude~\shortcite{gough78}.
In this limit we know that
\begin{align}
	\left.\nabla\times\left(\frac{1}{\rho}\nabla\cdot\mathbfss{T}\right)\right|_{\Omega} = 0.
	\label{eq:stressNoRot}
\end{align}
In the slowly-rotating limit then we can compute the contribution of the stress to the vorticity balance by only consider how rotation perturbs $\mathbfss{T}$.
Using symmetry arguments about the way different perturbations behave under reflections we estimate these perturabtions to be (Appendix~\ref{subsec:slow})
\begin{align}
	\nabla\times\left(\frac{1}{\rho}\nabla\cdot\mathbfss{T}\right)_r &\approx \left(\lambda + \frac{h}{r}\right)|N|\left(\Omega+|R\nabla\Omega|\right)\left(1 + \xi + \frac{u}{h|N|}\right),\\
	\nabla\times\left(\frac{1}{\rho}\nabla\cdot\mathbfss{T}\right)_\theta &\approx |N|\left(\Omega+|R\nabla\Omega|\right)\left(1 + \xi + \frac{u}{h|N|}\right)\\
\intertext{and}
	\nabla\times\left(\frac{1}{\rho}\nabla\cdot\mathbfss{T}\right)_\phi &\approx |N|^2 \left(\xi + \frac{u}{h} + \frac{\Omega^2}{|N|^2} + \frac{\Omega|R\nabla\Omega|}{|N|^2}\right),
\end{align}
Here we have assumed that the stress is analytic about $\Omega=0$ and that the leading order terms are always the lowest-order ones which are allowed by the symmetries of the problem.

\subsection{Advective Terms ($\Omega \ll |N|$)}

The advective terms in the vorticity equation are estimated in Appendix~\ref{appen:tab2} following the approximations introduced in Appendix~\ref{sec:baro}.

\subsection{Baroclinicity ($\Omega \ll |N|$)}

To proceed further we must determine the baroclinicity.
We do this by analyzing equation~\eqref{eq:heat1}, accounting for the fact that the diffusivity tensor is no longer isotropic in the presence of rotation.
We again employ symmetry arguments as well as the approximations of section~\ref{sec:merid}.
The result is equation~\eqref{eq:u_slow_2}
\begin{align}
		u + \frac{h}{|N|}\left(\left(|R\nabla\Omega| + \Omega\right)^2+ |N|^2 \xi\right) &\approx0,
		\label{eq:u_slow}
\end{align}
which serves to relate the meridional circulation, differential rotation, and baroclinicity through the condition of thermal equilibrium.

\subsection{Results ($\Omega \ll |N|$)}
\label{sec:slowres}

\begin{table}
\centering
\label{tab:sum1}
\begin{tabular}{lll}
Term & Magnitude\\
\hline
Meridional~\eqref{eq:vortm1} &\\
\hline\\
$\boldsymbol{\omega}_m\cdot\nabla\boldsymbol{u}$ & $\frac{u}{h}(\Omega + |R\nabla\Omega|)$\\
$\boldsymbol{\omega}_m \boldsymbol{u}\cdot\nabla\ln \rho$ & $\frac{u}{h}(\Omega + |R\nabla\Omega|) \left(\lambda + \frac{h}{r}\right)$\\
$\boldsymbol{u}\cdot\nabla\boldsymbol{\omega}_m$ & $\frac{u}{h}(\Omega + |R\nabla\Omega|)$\\
$\left(\nabla\times\left(\frac{1}{\rho}\nabla\cdot\mathbfss{T}\right)\right)_m$ & $|N|\left(\Omega+|R\nabla\Omega|\right)\left(1 + \xi + \frac{u}{h|N|}\right)$\\
\hline
Azimuthal~\eqref{eq:vortp1} & \\
\hline
$R^{-1}\omega_\phi u_R$ & $\frac{u^2}{hR}$\\
$\omega_\phi \boldsymbol{u}\cdot\nabla\ln\rho$ & $\frac{u^2}{h^2} \left(\lambda + \frac{h}{r}\right)$\\
$\boldsymbol{u}\cdot\nabla\omega_\phi$ & $\frac{u^2}{h^2}$\\
$\boldsymbol{\omega}\cdot\nabla(\Omega R)$ & $\Omega|R\nabla\Omega|$\\
$\Omega\omega_R$ & $\Omega|R\nabla\Omega|$\\
$\rho^{-2}(\nabla P\times\nabla \rho)_\phi$ & $\xi |N|^2$\\
$\left(\nabla\times\left(\frac{1}{\rho}\nabla\cdot\mathbfss{T}\right)\right)_\phi$ & $|N|^2 \left(\xi + \frac{u}{h|N|} + \frac{\Omega^2}{|N|^2} + \frac{\Omega|R\nabla\Omega|}{|N|^2}\right)$\\
\hline
\end{tabular}
\caption{The magnitudes of the terms in equations~\eqref{eq:vortm1} and~\eqref{eq:vortp1} are summarised here. Factors of order unity have been dropped for simplicity. See appendix~\ref{appen:tab2} for derivations of each of these.}
\end{table}

We now have enough information to estimate the magnitude of each term in the heat equation~\eqref{eq:heat0} and the meridional and azimuthal vorticity equations (\ref{eq:vortm1} and~\ref{eq:vortp1}).
Putting it all together and dropping sub-dominant terms these equations may be written in terms of the magnitudes of their terms as
\begin{align}
	0 &= \frac{u}{h}|N| + \xi |N|^2 + \left(\Omega + |R\nabla\Omega|\right)^2\\
	\label{eq:vortm3}
	0 &= \frac{u}{h}\left(\Omega+|R\nabla\Omega|\right) + |N|\left(\Omega+|R\nabla\Omega|\right)\left(1 + \xi + \frac{u}{h|N|}\right)\\
\intertext{and}
	\label{eq:vortp3}
	0 &= \frac{u^2}{h^2} + \Omega|R\nabla\Omega| + |N|^2 \xi + \Omega^2.
\end{align}

Consider starting with $\Omega=0$ and gradually spinning the system up.
The first equation indicates that the perturbation to the heat transport is of order $\Omega^2$.
This may be balanced either by a meridional circulation scaling similarly, baroclinicity scaling similarly or rotational shear scaling as $\Omega$.
Suppose for the moment that all three of these are realized.
Then $1 \gg \Omega/|N| \gg \xi,|N| \gg u/h$, so the meridional component reduces to
\begin{align}
	0 &= |N|\left(\Omega+|R\nabla\Omega|\right),
\end{align}
which is indeed solved asymptotically by $|R\nabla\Omega| \approx \Omega$.
Finally in the azimuthal vorticity equation with $u/h|N| \approx \Omega^2/|N|^2$ we see that the meridional circulation drops out, leaving
\begin{align}
	0 &= \Omega|R\nabla\Omega| + |N|^2 \xi + \Omega^2,
\end{align}
which is consistent with $\xi \approx \Omega^2/|N|^2$ and $|R\nabla\Omega| \approx \Omega$.

An exhaustive search of asymptotic solutions as described in section~\ref{sec:stiff} reveals three other consistent choices of scalings.
Each of these has $|R\nabla\Omega| \approx \Omega$, and they all have one of $\xi \approx \Omega^2/|N|^2$ or  $u/h|N| \approx \Omega^2/|N|^2$.
The other scales as a higher power of $\Omega/|N|$.
That is, they all require the same shear but allow for tuning such that equation~\eqref{eq:heat0} is satisfied with just one of the meridional circulation or baroclinicity.
Without imposing further assumptions we therefore have
\begin{align}
	|\nabla\Omega| &\approx \frac{\Omega}{R},\\
	u_\theta &\la u \approx h|N|\left(\frac{\Omega^2}{|N|^2}\right),\\
	\xi &\la \frac{\Omega^2}{|N|^2}.
\end{align}
With equation~\eqref{eq:upq2} we then obtain
\begin{align}
	u_r &\approx \frac{h}{r} u_\theta,
\end{align}
and with equation~\eqref{eq:lambdaxi} we find
\begin{align}
	\lambda &\la \frac{\Omega^2 h}{g}.
\end{align}
As discussed in Section~\ref{sec:stiff} we suggest that both the meridional circulation and baroclinicity scale similarly so that actually
\begin{align}
	\label{eq:resSlow}
	|\nabla\Omega| &\approx \frac{\Omega}{R},\\
	u_r &\approx \frac{h}{r} u_\theta,\\
	u_\theta &\approx u \approx h|N|\left(\frac{\Omega^2}{|N|^2}\right),\\
	\xi &\approx \frac{\Omega^2}{|N|^2}
\intertext{and}
	\lambda &\approx \frac{\Omega^2 h}{g}.	
\end{align}

%%%---------- close: slow_rot

%%%%%%%%%%% jump to rapid_hydro
%%%---------- open: rapid_hydro

\section{Hydrodynamic Rapid Rotation}
\label{sec:fast_hydro}

In this section we examine the case of rapid rotation with respect to the \brvs\ frequency, such that $\Omega \gg |N|$.
We neglect magnetic fields, which we shall consider in the next section.
However it is important to be careful because the relationship between the \brvs\ frequency and various convective quantities is altered in the limit of rapid rotation so, to be clear, we take $|N|$ to be the actually realized \brvs\ frequency and $|N|_0$ to be what the \brvs\ frequency would be were the rotation slow and all else held constant.
In this notation, the rapid rotation limit is that in which $\Omega \gg |N|_0$.
We are not interested in arbitrarily large rotation.
In particular the system must remain primarily pressure supported and so we also require that $\Omega \ll \sqrt{g/r}$.
The combination of these two limits is only sensible because, unlike in radiative zones, in convection zones $|N|$ may be significantly smaller than $\sqrt{g r^{-1}}$.

\subsection{Convection Speed}

The Coriolis effect stabilizes motion perpendicular to the rotation axis.
This means that in a rapidly rotating system fewer modes are unstable to convection, and those which remain unstable likely saturate at a smaller amplitude.

Many attempts have been made to estimate the strength of this effect through both closure models and numerical simulations, producing a variety of results including $\varv_c\propto \Omega^{-1/2}$~\citep{1990JFM...219..215B,2011Icar..211.1258S} and $\varv_c \propto \Omega^{-1}$~\citep{1979GApFD..12..139S,2014ApJ...791...13B}.
In section~\ref{sec:heat_flux} we shall adopt the latter of these scaling relations, motivated by the recent convincing suite of numerical simulations by~\citet{2020MNRAS.493.5233C}\footnote{\citet{jermyn} obtained $\varv_c \propto \Omega^{-1/2}$. The difference between our calculations there and those of~\citet{1979GApFD..12..139S} is that we did not impose the lower bound on the vertical wavenumber which they do. Such a lower bound is physically motivated for stars by the finite scale height, so we favour their scaling here.}.
However, at this stage we do not need to pick a scaling and deferring this decision keeps our analysis more general so we simply write
\begin{align}
	\varv_c \approx h |N| k,
	\label{eq:fastVC}
\end{align}
where $k(\Omega/|N|)$ is a continuous function which is order unity for $\Omega < |N|$ and which decreases asymptotically at least as fast as $(\Omega/|N|)^{-1/2}$ and no faster than $(\Omega/|N|)^{-1}$.

\subsection{Stress ($\Omega \gg |N|$)}
\label{sec:RapidHydroStress}

In Appendix~\ref{subsec:rapid} we estimate the magnitude of the shear using our estimates of the convective velocity and \brvs\ frequency.
In the limit of rapid rotation spherical symmetry is strongly broken, so it is no longer a guide as to how different components of the stress scale.
We assume that, because there is no symmetry protection, every component of the stress feels the baroclinicity and differential rotation and shear at linear order.
We thus obtain (equations~\ref{eq:hr1},~\ref{eq:hr2} and~\ref{eq:hr3})
\begin{align}
	\label{eq:hydro_rapid_1}
		\nabla\times\left(\frac{1}{\rho}\nabla\cdot\mathbfss{T}\right)_r &\approx \left(\left(\xi |N| + \frac{u}{h}+|R\nabla\Omega|\right)\left(|N|+|R\nabla\Omega|\right) + |N|^2\right) k^2\nonumber\\
		&\times \left(\frac{h}{r} + \frac{\lambda}{\bar{\lambda}}\right),
\end{align}
where $k$ is defined by equation~\eqref{eq:fastVC}.
Likewise, equations~\eqref{eq:FTot},~\eqref{eq:FTwt} and~\eqref{eq:viscFastT} give
\begin{align}
	\label{eq:hydro_rapid_2}
		\nabla\times\left(\frac{1}{\rho}\nabla\cdot\mathbfss{T}\right)_\theta &\approx \left(\left(\xi |N| + \frac{u}{h}+|R\nabla\Omega|\right)\left(|N|+|R\nabla\Omega|\right) + |N|^2\right)k^2,
\end{align}
and finally equations~\eqref{eq:FTop},~\eqref{eq:FTwp} and~\eqref{eq:viscFastP} produce
\begin{align}
	\label{eq:hydro_rapid_3}
		\nabla\times\left(\frac{1}{\rho}\nabla\cdot\mathbfss{T}\right)_\phi &\approx\left(\left(\xi |N| + \frac{u}{h}+|R\nabla\Omega|\right)\left(|N| + |R\nabla\Omega|\right) + |N|^2\right)k^2.
\end{align}

\subsection{Advective Terms ($\Omega \gg |N|$)}
\label{sec:RapidHydroAdv}

We now evaluate the terms which depend on $\boldsymbol{\varv}$ and its derivatives but which are not a part of the stress.
None of our analysis to determine the terms involving $\boldsymbol{u}$ in the meridional vorticity equation depended on the rotation being slow, so those may be found in appendix~\ref{appen:tab2}.
Only two terms couple the rotation to the $\boldsymbol{e}_z$ component of the differential rotation, so
\begin{align}
	|\boldsymbol{\omega}\cdot\nabla(\Omega R)| &\approx |R \Omega \partial_z \Omega|\\
	\intertext{and}
	|\Omega\omega_R| &\approx R \Omega |\partial_z \Omega|.
\end{align}
This is the only place where we encounter an intrinsic directional preference in the coupling of differential rotation to the vorticity equation.

\subsection{Baroclinicity ($\Omega \gg |N|$)}
\label{sec:RapidHydroBaro}

As well as its contribution to the stress tensor, the baroclinicity $\xi$ enters into the vorticity balance both in relation to the scale of the meridional circulation and through the thermal wind term.
So we must estimate $\xi$.
Once more we cannot rely on symmetry arguments so we make this estimate assuming that all symmetries are maximally broken.
In Appendix~\ref{subsec:pert_fast} we thus find that (equation~\ref{eq:u_MHD11})
\begin{align}
	\left(\xi + k\right) \frac{u}{h|N|} + k \left(1 + \xi + \frac{|R\nabla\Omega|}{\Omega}\right) \approx 0,
	\label{eq:u_hydro1}
\end{align}
which is a form of the heat equation~\eqref{eq:heat1} in this limit.

\subsection{Results ($\Omega \gg |N|$)}

\begin{table}
\centering
\begin{tabular}{lll}
Term & Magnitude\\
\hline
Meridional~\eqref{eq:vortm1} &\\
\hline\\
$\boldsymbol{\omega}_m\cdot\nabla_m\boldsymbol{u}$ & $\frac{u}{h}(\Omega + |R\nabla\Omega|)$\\
$\boldsymbol{\omega}_m \boldsymbol{u}\cdot\nabla\ln \rho$ & $\frac{u}{h}(\Omega + |R\nabla\Omega|) \left(\frac{h}{r} + \frac{\lambda}{\bar{\lambda}}\right)$\\
$\boldsymbol{u}\cdot\nabla\boldsymbol{\omega}_m$ & $\frac{u}{h}(\Omega + |R\nabla\Omega|)$\\
$\left(\nabla\times\left(\frac{1}{\rho}\nabla\cdot\mathbfss{T}\right)\right)_m$ & $k^2 |N| \left(|N| + \xi |N| + \frac{u}{h}+|R\nabla\Omega|\right)$\\
\hline
Azimuthal~\eqref{eq:vortp1} & \\
\hline
$R^{-1}\omega_\phi u_R$ & $\frac{u^2}{hR}$\\
$\omega_\phi \boldsymbol{u}\cdot\nabla\ln\rho$ & $\frac{u^2}{h^2} \left(\frac{h}{r} + \frac{\lambda}{\bar{\lambda}}\right)$\\
$\boldsymbol{u}\cdot\nabla\omega_\phi$ & $\frac{u^2}{h^2}$\\
$\boldsymbol{\omega}\cdot\nabla(\Omega R)$ & $\Omega|R\nabla\Omega|$\\
$\Omega\omega_R$ & $\Omega|R\nabla\Omega|$\\
$\rho^{-2}(\nabla P\times\nabla \rho)_\phi$ & $\xi |N|^2$\\
$\left(\nabla\times\left(\frac{1}{\rho}\nabla\cdot\mathbfss{T}\right)\right)_\phi$ & $k^2 |N| \left(|N| + \xi |N| + \frac{u}{h}+|R\nabla\Omega|\right)$\\
\hline
\end{tabular}
\caption{The magnitudes of the terms in equations~\eqref{eq:vortm1} and~\eqref{eq:vortp1} are summarized here. Factors of order unity have been dropped for simplicity.}
\label{tab:sum4}
\end{table}

Our results thus far are summarized in Table~\ref{tab:sum4}.
To proceed we must balance equations~\eqref{eq:heat0},~\eqref{eq:vortm1} and~\eqref{eq:vortp1} for $u$, $\xi$ and $R\nabla\Omega$.
We do this using the methods of section~\ref{sec:stiff}, beginning with a heuristic argument and ending with an exhaustive search of the possibilities.

The heat equation~\eqref{eq:heat0} is satisfied by having either $|R\nabla\Omega| \approx \Omega$, $u/h|N| \approx k/\xi$ or $\xi \approx 1$.
We argued in section~\ref{sec:stiff} that the first of these is not allowed in the limit of rapid rotation, so one of the other two must hold.

In the meridional vorticity equation $u$ enters one order higher in $\Omega/|N|$ than all other terms.
This suggests that it must asymptotically decrease like $k^2 |N|/\Omega$ in order for that equation to be balanced.
This is smaller than required to satisfy the heat equation~\eqref{eq:heat0} with the meridional circulation dominant, so we find $\xi \approx 1$.

In the azimuthal vorticity balance the term $\Omega|R\nabla\Omega|$ is the Taylor-Proudman term~\citep{1897RSPTA.189..201H}.
As we show in appendix~\ref{appen:tab2} this piece is actually $\Omega|R\partial_z \Omega|$.
Other occurrences of the differential rotation in this equation are sensitive to both components of the differential rotation.
In this equation then, as in the other two, the shear is comparable to the dominant terms we have identified if $|R\nabla\Omega| \approx |N|$ and $|R\partial_z \Omega| \approx |N|^2/\Omega$.

An exhaustive search of the consistent scalings with the extremal options of $k\propto \Omega^{-1/2}$ and $k\propto \Omega^{-1}$ reveals several other solutions.
Each of these has $\xi$ of order unity, not scaling with $\Omega/|N|$, and has $|R\partial_z\Omega| \approx |N|^2/\Omega$.
The differences permit various tradeoffs of tuning $u$ and the other component of the shear.
Assuming as before that no tuning occurs, so that each of these is as large as it can be
\begin{align}
	\xi &\approx 1,\\
\label{eq:hydroDR}
	|R\nabla\Omega| &\approx |N|,\\
\label{eq:hydroDZ}
	|R\partial_z\Omega| &\approx \frac{|N|^2}{\Omega}\\
\intertext{and}
	u &\approx h |N| \frac{|N|}{\Omega} k^2.
\end{align}
Note that the scaling of $|R\nabla\Omega|$ is consistent with the findings of~\citet{2011Icar..211.1258S} in what they term the asymptotic regime, and corresponds to our rapidly rotating regime when all microscopic diffusivities vanish.

%%%---------- close: rapid_hydro

%%%%%%%%%%% jump to rapid_rot
%%%---------- open: rapid_rot

\section{Magnetic Rapid Rotation}
\label{sec:fast}

We now repeat the analysis of the previous section including the effects of magnetic fields.

\subsection{Magnetic Fields ($\Omega \gg |N|$)}
\label{sec:mag_field}

If the magnetic diffusivity is vanishingly small, then the growth of the dynamo is limited only by the fact that above equipartition the field begins to quench convection~\citep{1989ApJ...342.1158M}.
This simple argument predicts
\begin{align}
	\varv_\mathrm{A} \approx h |N|.
	\label{eq:vA_hN0}
\end{align}
However, both analytical arguments and numerical simulations show considerable dispersion in the scaling of magnetic field strength with buoyancy and rotation.
\citet{1979GApFD..12..139S} predicted that with fixed heat flux the magnetic field scales as $\Omega^{1/4}$ from analytic growth-rate arguments.
Several recent arguments suggest similar scaling laws~\citep{2002Icar..157..426S,2017JFM...813..558A}, though others obtain $B \propto \Omega^{0}$~\citep{2013GeoJI.195...67D}.
In numerical simulations, the field strength has been found to scale as $\Omega^{-0.02}$~\citep{doi:10.1111/j.1365-246X.2006.03009.x}\footnote{See their equation (33).}, $\Omega^{-0.11}$~\citep{2013Icar..225..185Y}\footnote{From their dipolar fit with no magnetic Prandtl number dependence.}, $\Omega^{0}$~\citep{2013Icar..225..185Y}\footnote{From their dipolar fit with magnetic Prandtl number dependence.}, and between $\Omega^{-0.02}$ and $\Omega^{0.19}$~\citep{2017JFM...813..558A}\footnote{See their figure 11(a). Their $\epsilon$ is proportional to $\Omega^{-3}$ and their $\lambda \propto B \Omega^{-1}$.}.

Given this uncertainty, we parameterize the scaling of the magnetic field instead by
\begin{align}
\varv_\mathrm{A} \approx h |N| q,
\label{eq:vA_hN}
\end{align}
where $q\left(\frac{\Omega}{|N|}\right)$ is a continuous function which is approximately constant when $\Omega \ll |N|$.
This form is consistent with observations that magnetic activity is principally a function of Rossby number ($|N|/\Omega$)~\citep{Lehtinen_2020}.
In all suggested scalings of which we are aware $q \ga k$ when $\Omega \gg |N|$, so the magnetic field becomes super-equipartion in this limit and the magnetic field energy exceeds the convective energy by a factor of $(q/k)^{-2}$.

\subsection{Modifications}

The arguments of sections~\ref{sec:RapidHydroStress} apply to this scenario with just one modification, namely that the scale of the turbulence is set by the \alf\ speed, which means that the stress is proportional to $q^2 |N|^2$ rather than $k^2 |N|^2$.
With this equations~\eqref{eq:hydro_rapid_1},~\eqref{eq:hydro_rapid_2} and~\eqref{eq:hydro_rapid_3} become
\begin{align}
		\nabla\times\left(\frac{1}{\rho}\nabla\cdot\mathbfss{T}\right)_r &\approx q^2 |N|^2\left(1 + \xi + \frac{|R\nabla\Omega|}{|N|} + \frac{u}{h|N|}\right)\\
		&\times \left(\frac{h}{r} + \frac{\lambda}{\bar{\lambda}}\right),\nonumber\\
		\nabla\times\left(\frac{1}{\rho}\nabla\cdot\mathbfss{T}\right)_\theta &\approx q^2 |N|^2\left(1 + \xi + \frac{|R\nabla\Omega|}{|N|} + \frac{u}{h|N|}\right)\\
\intertext{and}
		\nabla\times\left(\frac{1}{\rho}\nabla\cdot\mathbfss{T}\right)_\phi &\approx q^2 |N|^2\left(1 + \xi + \frac{|R\nabla\Omega|}{|N|} + \frac{u}{h|N|}\right).
\end{align}

\subsection{Results ($\Omega \gg |N|$)}

The results of sections~\ref{sec:RapidHydroAdv} and section~\ref{sec:RapidHydroBaro} still apply.
The same heuristic argument applies as in section~\ref{sec:fast_hydro}, except that now the stress is a factor of $(q/k)^2$ larger in the meridional vorticity equation, so the meridional circulation $u$ is enhanced by a factor of $(q/k)^{2}$.
Note that as before the meridional circulation is limited most strongly by the meridional vorticity equation and not the heat equation, so the change in the heat equation from the hydrodynamic case does not enter into our calculations.
The remainder of the argument is unchanged, so
\begin{align}
	\xi &\approx 1,\\
\label{eq:rapid_MHD_DR}
	|R\nabla\Omega| &\approx |N|,\\
	|R\partial_z\Omega| &\approx \frac{|N|^2}{\Omega}\\
\intertext{and}
	u &\approx h |N| q^2 \frac{|N|}{\Omega}.
\end{align}
Once more we performed an exhaustive search for other consistent scaling relations, this time with all combinations of $k \propto \Omega^{-1/2}$, $k \propto \Omega^{-1}$, $q \propto \Omega^{-1/2}$ and $q \propto \Omega^0$.
All of the other consistent relations have $\xi \approx 1$.
The result above is recovered by assuming, as before, that no tuning occurs such that each of these achieves their greatest allowed value.

Note that with the total shear scaling as $|R\nabla\Omega| \approx |N|$ and the stress scaling as $\Omega |R\nabla\Omega|$, the overall stress is similar to that of the saturated magnetorotational instability (MRI)~\citep{Wheeler_2015}.
%%%---------- close: rapid_rot

%%%%%%%%%%% jump to heat_flux
%%%---------- open: heat_flux
\section{Heat Flux}
\label{sec:heat_flux}

We must account for the fact that the Coriolis effect arrests convective motions, so that at fixed $|N|$ increasing $\Omega$ decreases the heat flux.
Equivalently, at fixed heat flux $|N|$ must increase with increasing $\Omega$.
If the rotation does not significantly affect the heat source in a star, or the outer boundary condition in a planet, $|N|$ must change to keep $F$ constant.

To estimate this effect when $\Omega \gg |N|$ we use
\begin{align}
	k \approx \frac{|N|}{\Omega}.
	\label{eq:k_Adrian}
\end{align}
This scaling was obtained by~\citet{1979GApFD..12..139S} through analytic closure arguments and has been observed in the extensive suites of simulations by~\citet{2014ApJ...791...13B} and~\citet{2020MNRAS.493.5233C}.

\subsection{Hydrodynamic}

The modulus of equation~\eqref{eq:QF} is
\begin{align}
	F = \mu^{-1}\rho c_p T |\mathbfss{Q}\cdot\nabla s|.
\end{align}
Assuming the inner product produces a factor of order unity we obtain
\begin{align}
	F \approx \mu^{-1}\rho c_p T |\mathbfss{Q}||\nabla s|.
\end{align}
We may approximate the diffusivity $\mathbfss{Q}$ by that of a random walk with velocity equal to the convective velocity and time-scale given by the \brvs\ frequency so that
\begin{align}
	F \approx \mu^{-1} \rho c_p T \frac{\varv_{\mathrm{c}}^2}{|N|} |\nabla s|.
\end{align}
Inserting equation~\eqref{eq:brvs} and using $\bar{\xi}\approx 1$ we find
\begin{align}
	F \approx \frac{\rho^2 c_p T}{\mu p}\varv_{\mathrm{c}}^2 h |N|.
\end{align}
With the ideal gas law this simplifies to
\begin{align}
	F \approx \rho \varv_{\mathrm{c}}^2 h |N|.
\end{align}
Inserting equation~\eqref{eq:fastVC} we find
\begin{align}
	F \approx \rho h^3 |N|^3 k^2.
\end{align}
With equation~\eqref{eq:k_Adrian} this becomes
\begin{align}
	F \approx \rho h^3 |N|^5 \Omega^{-2}.
\end{align}
So for fixed heat flux
\begin{align}
\label{eq:hydroN}
	|N| &\approx |N|_0 \left(\frac{\Omega}{|N|_0}\right)^{2/5}\\
\intertext{and}
\label{eq:hydroVC}
	\varv_c &\approx h |N|_0 \left(\frac{\Omega}{|N|_0}\right)^{-1/5},
\end{align}
consistent with what is seen in simulations~\citep{2016JFM...808..690G}.

\subsection{Magnetohydrodynamic}

The magnetic case is more complicated.
A consistent finding in both numerical simulations~\citep{10.1093/gji/ggt083} and analytic arguments~\citep{doi:10.1111/j.1365-246X.2006.03009.x} is that
\begin{align}
	F \approx \rho \varv_c \varv_\mathrm{A}^2,
	\label{eq:varvAf}
\end{align}
which can be interpreted as the heat flux coming about primarily from advection of magnetic energy rather than entropy.
In addition to simulations, equation~\eqref{eq:varvAf} also agrees with observations of Jupiter~\citep{doi:10.1111/j.1365-246X.2006.03009.x}.
There are more significant differences when compared with the observed surface magnetic field of Saturn but there is reason to believe that this is not reflective of its interior~\citep{1982GApFD..21..113S}.

Next note that from equation~\eqref{eq:vA_hN} the magnetic acceleration is of order $h|N|^2 q^2$.
By contrast the Coriolis effect acting on the convection speed given by equation~\eqref{eq:fastVC} produces an acceleration of order $h |N| \Omega k$.
So long as
\begin{align}
	q^2 \la k \frac{\Omega}{|N|}
	\label{eq:qk}
\end{align}
the magnetic force is no larger than the Coriolis force, so the convection speed is limited by rotation and we expect equation~\eqref{eq:fastVC} to remain valid.
Equation~\eqref{eq:qk} holds for most of the scalings we are aware of in the literature, so we shall assume that $k$ is unchanged.

We now let $q = 1$, and shall show that this generates the same scalings as those found by~\citet{1979GApFD..12..139S} and~\citet{2002Icar..157..426S}.
Inserting $q=1$ into equation~\eqref{eq:vA_hN} we expect the ratio of magnetic to kinetic energy to scale like
\begin{align}
\frac{\varv_\mathrm{A}^2}{\varv_c^2} \approx \frac{\Omega^2}{|N|^2}.
\end{align}
This is consistent with simulations by~\citet{2016ApJ...829...92A} and~\citet{2019arXiv190400225C}, which show that the ratio of magnetic to kinetic energy scales as the inverse Rossby number
\begin{align}
\frac{\varv_\mathrm{A}^2}{\varv_c^2}\approx \mathrm{Ro}^{-1} \approx \frac{\Omega h}{\varv_c} \approx \frac{\Omega^2}{|N|^2},
\label{eq:rosmag}
\end{align}
where we have inserted equation~\eqref{eq:k_Adrian} in the last step.

We proceed with equation~\eqref{eq:k_Adrian}.
Combining this with equations~\eqref{eq:varvAf},~\eqref{eq:fastVC} and~\eqref{eq:varvAf} we find
\begin{align}
	F&\approx\rho \varv_{\mathrm{c}} \varv_{\mathrm{A}}^2 \approx \rho h^3 |N|^3 k \approx \rho h^3 |N|^4 \Omega^{-1},
	\label{eq:FA}
\end{align}
so at fixed heat flux
\begin{align}
\label{eq:magN}
	|N| &\approx |N|_0 \left(\frac{\Omega}{|N|_0}\right)^{1/4},\\
\label{eq:magVC}
	\varv_c &\approx h |N|_0 \left(\frac{\Omega}{|N|_0}\right)^{-1/2},\\
	\intertext{and}
\label{eq:magA}
	\varv_\mathrm{A} &\approx h |N|_0 \left(\frac{\Omega}{|N|_0}\right)^{1/4}.
\end{align}
This scaling for $\varv_{\mathrm{A}}$ was predicted by~\citet{1979GApFD..12..139S} and more recently by~\citet{2002Icar..157..426S}, who further produced the scalings \brvs\ frequency and convection speed in equations~\eqref{eq:magN} and~\eqref{eq:magVC}.
There is significant scatter in the corresponding scalings produced by simulations, but generally they suggest a weaker scaling for $\varv_c$ than what we find here.
For instance the convection speed at constant heat flux is seen to scale as $\Omega^{-0.23}$ to $\Omega^{-0.29}$~\citep{doi:10.1111/j.1365-246X.2006.03009.x}\footnote{See their equations (30) and (31).}, $\Omega^{-0.41}$~\citep{2013Icar..225..185Y}\footnote{From their multipolar fit with magnetic Prandtl number dependence.} and $\Omega^{-0.32}$ to $\Omega^{-0.47}$~\citep{2017JFM...813..558A}\footnote{See their figure 11(a). Their $\epsilon$ is proportional to $\Omega^{-3}$.}.
Similarly the scalings of the magnetic field which we discuss in section~\ref{sec:mag_field} are generally a bit weaker than that in equation~\eqref{eq:magA}, so it is possible that $q$ ought to mildly decrease with $\Omega/|N|$.

To bracket the possibilities note that a scaling of the form $q\propto \Omega^{\beta}$ for $\beta \la -1/2$ produces an equal amount of tension in the opposite direction, and is ruled out by the scaling of the ratio of magnetic to kinetic energy~\citet{2019arXiv190400225C}.
So it may be that the truth lies between these.
Given the precision of current observations of differential rotation the difference between these choices of $q$ is small, producing relative shear scaling like either $\Omega^{-3/4}$ or $\Omega^{-3/5}$.
If future numerical simulations pin down the scaling of magnetic field strength more narrowly we can always revisit this scaling, but for now we take $q=1$ as the simpler option.

Note that with this choice we may combine equations~\eqref{eq:varvAf} and~\eqref{eq:QF} to find
\begin{align}
	 \mathsf{Q} |\nabla s| \approx \frac{\rho}{P} \varv_{\mathrm{c}} \varv_{\mathrm{A}}^2.
\end{align}
Rearranging equation~\eqref{eq:brvs}, inserting equation~\eqref{eq:h} and dropping factors of order unity yields
\begin{align}
	|\nabla s| \approx \frac{\rho}{P} h |N|^2.
\end{align}
Inserting this into the previous relation we obtain
\begin{align}
	 \mathsf{Q}  \approx \frac{\varv_{\mathrm{c}} \varv_{\mathrm{A}}^2}{h|N|^2}
\end{align}
and recalling that $\varv_\mathrm{A} \approx h|N|$ we find
\begin{align}
	 \mathsf{Q}  \approx \varv_{\mathrm{c}} h,
	 \label{eq:QMHD}
\end{align}
which is the same diffusivity we used in the hydrodynamic case.

%%%---------- close: heat_flux

%%%%%%%%%%% jump to limitations
%%%---------- open: limitations
\section{Limitations}
\label{sec:limitations}

To reiterate from section~\ref{sec:assump}, we have made the following assumptions.
\begin{enumerate}
	\item Dimensionless factors arising from geometry are of order unity unless symmetries require them to be otherwise.
	\item All external perturbing forces, such as tides or external heating, are negligible in the regions of interest.
	\item The material is non-degenerate, compressible and not radiation-dominated.
	\item All microscopic (i.e. non-turbulent) diffusivities are negligible, such that:
	\begin{enumerate}
	\item convection is efficient, so the gas is nearly isentropic,
	\item the Reynolds and Rayleigh numbers are much larger than critical, and
	\item magnetohydrodynamical processes are ideal.
	\end{enumerate}
	\item The system is axisymmetric in a time-averaged sense.
	\item Convection is subsonic.
	\item The system is chemically homogeneous.
\end{enumerate}
We now consider each of these assumptions and explain how they limit our results.

We have already discussed geometric factors extensively at various points.
The main limitation they introduce is that we cannot easily incorporate further information about boundary conditions or the scale of the convection zone.
For instance, we cannot readily predict what ought to happen in a convecting shell surrounding a differentially-rotating sphere.
Near the sphere boundary effects dominate and our theory is inapplicable.

The next assumption is that perturbing forces such as tides or external sources of heating may be neglected.
In single systems this is valid but in binary or planetary systems it may not be.
In order for tides to be relevant the angular momentum transport they induce must be at least of order the steady state flux which is transported by the various terms which balance in the vorticity equation.
Likewise, in order for heating to be relevant, it must be at least of the order that arises from the rotational perturbations to the equation of thermal equilibrium.
In both cases our assumption is unlikely to be violated for stars but could fail in, for instance, hot Jupiter systems where the heat flux owing either to tides or to insolation may exceed that emerging from the centre~\citep{2017MNRAS.469.1768J}.
Similarly in a highly eccentric low-mass binary system the instantaneous tidal torque could be significant relative to convective angular momentum flux.
Such scenarios are rare but likely exist.

Our third assumption enters the analysis only insofar as it allows us to use an ideal gas-type equation of state.
In particular, in several places, we have used the fact that the pressure depends on both temperature and density.
Removing the dependence on temperature changes the structure of these arguments significantly and so we have simply ignored such cases.
This means that our analysis cannot be applied robustly to compact objects, rocky or otherwise solid bodies, or to degenerate planetary cores.
In such systems though the microscopic viscosity may be quite large and convection may not be fully developed and so we would need to exclude them anyway.
We further cannot apply our results to radiation-dominated regions of massive stars.

The fourth assumption is principally one of convenience: by neglecting microscopic diffusivities, we achieve significant simplifications of the equations of thermal and vorticity equilibrium.
Indeed,  in systems for which the third assumption holds the true momentum diffusivity is expected to be extremely small~\citep{1956pfig.book.....S}.
On the other hand the thermal diffusivity may not be small and near a radiative-convective boundary this assumption definitely fails.
Nevertheless, for systems in which a convecting region is large enough to matter for the rotation of the system, we expect it to also be large enough that such boundaries do not dominate its dynamics.
In effect this is an extension of the assumption that geometry, and hence boundary effects, are not too important.

Related to this, there is one place in which the microscopic thermal diffusivity cannot be neglected, namely near the limit of breakup rotation.
As we have mentioned, the scaling laws we have derived do not hold all the way to the breakup velocity.
We have shown that this is because a system which does not reconfigure to follow Keplerian rotation cannot continue to convect as the rotation approaches breakup.
However the way in which convection is disrupted is by reducing the effective gravity such that a radiative temperature gradient may be convectively stable.
This requires a finite radiative gradient and thence that the diffusivity does not vanish.
It may be arbitrarily small because reducing the diffusivity just shifts the rotation rate at which convection ceases closer to the breakup rate but for any non-zero diffusivity there is a rotation rate at which convection fails.
This may seem like a purely technical point but it is important to note because it precludes smoothly connecting the rapidly-rotating convecting solution to that of a Keplerian disc.
Notably this points to one of the key open problems in understanding heat transport in W~UMa\footnote{i.e. low-mass contact binary} systems~\citep{doi:10.1111/j.1365-2966.2004.07761.x}, namely that there must be regions in which convection fails because the effective gravity vanishes.

Along similar lines, in Jupiter the microscopic conductivity changes dramatically at the depths at which ionization occurs.
This could result in a transition from hydrodynamic to magnetohydrodynamic scaling.
Near the transition region the behavior could be more complex than in either limit.

The fifth assumption, that of axisymmetry, is a strong one.
It is responsible for a myriad of simplifications in our equations and, in particular controls, the orders of various perturbations which are protected by this symmetry.
As a result any phenomena which break this symmetry may introduce new modes of heat and momentum transport which violate our calculations by an amount which is proportional to the symmetry breaking.
This is a concern for systems which exhibit tides or non-axisymmetric external sources of heating.
In many cases these effects are either very low in amplitude, as in a long-period binary, or very high in frequency, as in a short-period binary.
In the former case they may be neglected owing to their amplitude, while in the latter they may produce no leading order effect because they are not well-matched frequency-wise to the turbulence.
This was noted by~\citet{1977ApJ...211..934G} in the context of tides, where at high frequencies relative to $|N|$ convection only couples weakly to the tidal potential.
Nevertheless, there are cases in which axisymmetry strongly fails, such as in W~UMa systems~\citep{doi:10.1111/j.1365-2966.2004.07761.x} and so this assumption is worth considering carefully when applying our results.

We have already discussed the assumption that convection is subsonic and so merely note that this would only result in incorrect scaling relations if the Mach number were to exceed unity by a factor which depended strongly on $\Omega/|N|$.
That is, if the Mach number exceeds unity by a factor of a few which is set by thermodynamic considerations our analysis is unchanged but if the Mach number can increase without bound as $\Omega \rightarrow 0$ or $\infty$ we have a problem because then we cannot bound the convection speed by thermodynamic considerations.
In fact we have shown that this is not the case because the convection speed is largely independent of $\Omega$ as $\Omega \rightarrow 0$ and decreases for fixed $|N|$ as $\Omega \rightarrow \infty$.
This assumption is therefore not one which we expect to be violated in any significant way.

Finally we must consider chemistry.
We have assumed everywhere that the system is chemically homogeneous.
This is actually quite likely because convection rapidly mixes chemical composition and so we do not expect to find substantial violation of this assumption unless, for instance, material is being injected into a convection zone at a rate comparable to the convective mass flux.
This is a rather exotic scenario though and, with a few notable exceptions\footnote{For example consider hot bottom burning in Asymptotic Giant Branch stars, planet injestion, He-flashes, etc.}, does not reflect a system which is undergoing evolution on secular or nuclear time-scales so we suffer no great loss by excluding it.
%%%---------- close: limitations

%%%%%%%%%%% jump to summary
%%%---------- open: summary

\section{Summary}
\label{sec:summ}

\begin{figure*}
\includegraphics[width=0.8\textwidth]{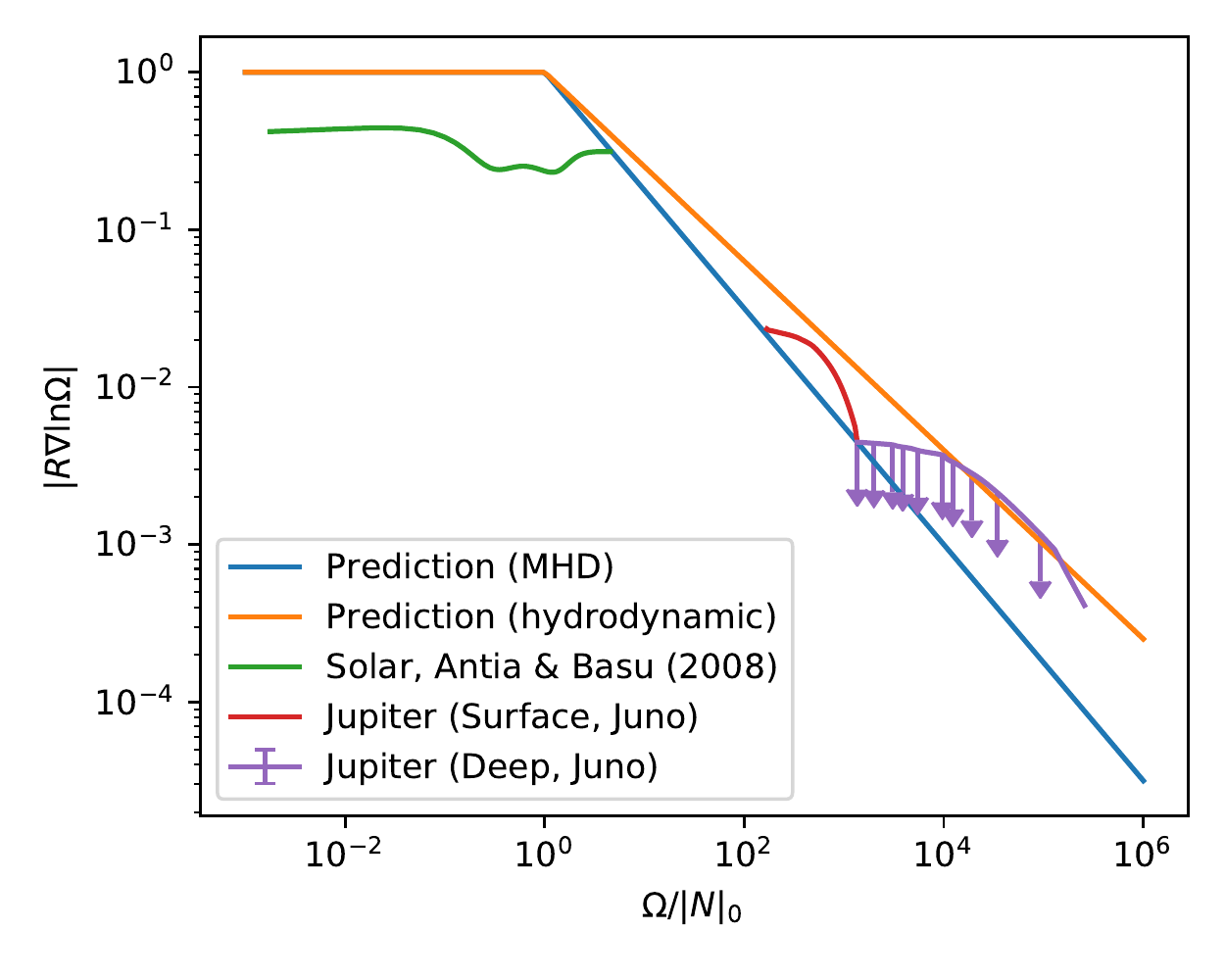}
\caption{Our prediction for the differential rotation given by Table~\ref{tab:summary} is shown as a function of $\Omega/|N|_0$ and normalised by $\Omega$. This is constant for $\Omega < |N|_0$ and scales like $(\Omega/|N|_0)^{-3/4}$ (MHD) and $(\Omega/|N|_0)^{-3/5}$ (hydrodynamic) for $\Omega > |N|_0$. Also shown are data for the Sun~\citep{2008ApJ...681..680A} and Jupiter~\citep{2018Natur.555..223K,2018Natur.555..227G}. The volume-weighted root-mean square shear in the Sun is shown at the average $\Omega/|N|_0$. {\emph{Juno}} measurements of the surface, at a depth of less than $3000\,\rm{km}$, shear are shown separately from {\emph{Juno}} upper limits on the shear deeper down. Details of the data analysis may be found in a companion paper where we focus on observational tests~\citep{PartII}.}
\label{fig:summary}
\end{figure*}

\begin{table*}
\caption{The scalings of the differential rotation, meridional circulation, baroclinicity, \brvs\ frequency, convective velocity, and the ratio of magnetic to kinetic energy are given for the three regimes of interest. Note that the latitudinal and spherical radial differential rotation are each formed of a mixture of the cylindrical vertical and radial differential rotation. Because the cylindrical radial shear is larger than the vertical shear, both spherical components of the differential rotation share the scaling of the former.}
\label{tab:summary0}
\begin{tabular}{llllll}
\hline
\hline
Case & $\frac{|R\nabla \Omega|}{\Omega}$ & $\frac{|R\partial_R \Omega|}{\Omega}$ & $\frac{|R\partial_z \Omega|}{\Omega}$ & $\frac{|r\partial_r \Omega|}{\Omega}$ &$\frac{|\partial_\theta \Omega|}{\Omega}$\\
\hline
Slow ($\Omega \ll |N|_0$) & $1$ & $1$ & $1$ & $1$ & $1$\\
Fast Hydro.($\Omega \gg |N|$) & $\left(\frac{\Omega}{|N|}\right)^{-1}$& $\left(\frac{\Omega}{|N|}\right)^{-1}$ & $\left(\frac{\Omega}{|N|}\right)^{-2}$& $\left(\frac{\Omega}{|N|}\right)^{-1}$& $\left(\frac{\Omega}{|N|}\right)^{-1}$\\
Fast MHD ($\Omega \gg |N|$) &$\left(\frac{\Omega}{|N|}\right)^{-1}$&$\left(\frac{\Omega}{|N|}\right)^{-1}$& $\left(\frac{\Omega}{|N|}\right)^{-2}$ &$\left(\frac{\Omega}{|N|}\right)^{-1}$&$\left(\frac{\Omega}{|N|}\right)^{-1}$\\
\hline
\hline
Case & $\frac{u_r}{h|N|}$ & $\frac{u_\theta}{h|N|}$ & $\xi$ &$\frac{\varv_{\rm c}}{h|N|}$ & $\frac{\varv_{\rm A}^2}{\varv_{\rm c}^2}$\\
\hline
Slow ($\Omega \ll |N|$) & $\frac{h}{r}\left(\frac{\Omega}{|N|}\right)^{2}$ & $\left(\frac{\Omega}{|N|}\right)^{2}$ & $\left(\frac{\Omega}{|N|}\right)^2$ & $1$ & $1$\\
Fast Hydro.($\Omega \gg |N|$) & $k^2 \frac{h}{r}\left(\frac{\Omega}{|N|}\right)^{-1}\approx \frac{h}{r}\left(\frac{\Omega}{|N|}\right)^{-3}$& $\left(\frac{\Omega}{|N|}\right)^{-1} k^2\approx\left(\frac{\Omega}{|N|}\right)^{-3}$ & $1$  & $k\approx \frac{|N|}{\Omega}$ & N/A\\
Fast MHD ($\Omega \gg |N|$) & $q^2 \frac{h}{r}\left(\frac{\Omega}{|N|}\right)^{-1}\approx \frac{h}{r}\left(\frac{\Omega}{|N|}\right)^{-1}$& $\left(\frac{\Omega}{|N|}\right)^{-1} q^2 \approx\left(\frac{\Omega}{|N|}\right)^{-1}$ & $1$  & $k\approx \frac{|N|}{\Omega}$ & $\frac{q^2}{k^2} \approx \frac{\Omega^2}{|N|^2}$\\
\hline
\hline
\end{tabular}
\end{table*}

\begin{table*}
\caption{The scalings of the differential rotation, meridional circulation, baroclinicity, \brvs\ frequency, convective velocity, and the ratio of magnetic to kinetic energy are given for the three regimes of interest in terms of the non-rotating \brvs\ frequency $|N|_0$ and expanding $q \approx 1$ and $k \approx |N|/\Omega$. Note that the latitudinal and spherical radial differential rotation are each formed of a mixture of the cylindrical vertical and radial differential rotation. Because the cylindrical radial shear is larger than the vertical shear, both spherical components of the differential rotation share the scaling of the former.}
\label{tab:summary}
\begin{tabular}{lllllll}
\hline
\hline
Case & $\frac{|R\nabla \Omega|}{\Omega}$ & $\frac{|R\partial_R \Omega|}{\Omega}$ & $\frac{|R\partial_z \Omega|}{\Omega}$ & $\frac{|r\partial_r \Omega|}{\Omega}$ &$\frac{|\partial_\theta \Omega|}{\Omega}$ &\\
\hline
Slow ($\Omega \ll |N|_0$) & $1$ & $1$ & $1$ & $1$ & $1$&\\
Fast Hydro.($\Omega \gg |N|_0$) & $\left(\frac{\Omega}{|N|_0}\right)^{-3/5}$& $\left(\frac{\Omega}{|N|_0}\right)^{-3/5}$ & $\left(\frac{\Omega}{|N|_0}\right)^{-6/5}$& $\left(\frac{\Omega}{|N|_0}\right)^{-3/5}$& $\left(\frac{\Omega}{|N|_0}\right)^{-3/5}$&\\
Fast MHD ($\Omega \gg |N|_0$) &$\left(\frac{\Omega}{|N|_0}\right)^{-3/4}$&$\left(\frac{\Omega}{|N|_0}\right)^{-3/4}$& $\left(\frac{\Omega}{|N|_0}\right)^{-3/2}$ &$\left(\frac{\Omega}{|N|_0}\right)^{-3/4}$&$\left(\frac{\Omega}{|N|_0}\right)^{-3/4}$&\\
\hline
\hline
Case & $\frac{u_r}{h|N|_0}$ & $\frac{u_\theta}{h|N|_0}$ & $\xi$ & $\frac{|N|}{|N|_0}$ &$\frac{\varv_{\rm c}}{h|N|_0}$ & $\frac{\varv_{\rm A}^2}{\varv_{\rm c}^2}$\\
\hline
Slow ($\Omega \ll |N|_0$) & $\frac{h}{r}\left(\frac{\Omega}{|N|_0}\right)^{2}$ & $\left(\frac{\Omega}{|N|_0}\right)^{2}$ & $\left(\frac{\Omega}{|N|}\right)^2$ & $1$ & $1$ & $1$\\
Fast Hydro.($\Omega \gg |N|_0$) & $\frac{h}{r}\left(\frac{\Omega}{|N|_0}\right)^{-7/5}$& $\left(\frac{\Omega}{|N|_0}\right)^{-7/5}$ & $1$ & $\left(\frac{\Omega}{|N|_0}\right)^{2/5}$ & $\left(\frac{\Omega}{|N|_0}\right)^{-1/5}$ & N/A\\
Fast MHD ($\Omega \gg |N|_0$) & $\frac{h}{r}\left(\frac{\Omega}{|N|_0}\right)^{-1/2}$ & $\left(\frac{\Omega}{|N|_0}\right)^{-1/2}$& $1$ & $\left(\frac{\Omega}{|N|_0}\right)^{1/4}$ & $\left(\frac{\Omega}{|N|_0}\right)^{-1/2}$ & $\left(\frac{\Omega}{|N|_0}\right)^{3/2}$\\
\hline
\hline
\end{tabular}
\end{table*}

In this work we studied differential rotation in both slowly and rapidly rotating convection zones in both the hydrodynamic and magnetohydrodynamic limits.
We obtained scaling laws for the differential rotation, the baroclinicity $\xi$ and the meridional circulation velocity.
These are summarized in Table~\ref{tab:summary0}.
In section~\ref{sec:heat_flux} we then incorporated scaling relations for the convection speed and magnetic field to obtain the scaling of the \brvs\ frequency with rotation rate at fixed heat flux.
This allows us to put our scaling relations in terms of the non-rotating \brvs\ frequency $|N|_0$ which is obtained from standard stellar evolution calculations.
These modified scaling relations are summarized in Table~\ref{tab:summary}.

Our findings for slowly rotating systems are consistent with the solar rotation profile~\citep{2015ApJ...813..114R} as well as the rotation profiles of other slowly-rotating systems~\citep{2009ApJ...702.1078B,2011A&A...531A.162K}.
Physically this results from a balance between turbulent viscosity and the $\Lambda$-effect in both equations, in agreement with arguments by~\citep{1989drsc.book.....R}.
In the azimuthal vorticity equation a comparable amount is also contributed by the thermal wind term.
This is in good agreement with the work of~\citet{2012MNRAS.420.2457B}, who find that thermal wind balance produces a good match to the solar rotation profile in the bulk of the solar convection zone.

The scaling we obtain for the differential rotation differs from that of~\citet{1998ApJ...494..691R} by a factor of $\Omega/|N|$.
This appears to be because they neglect both the thermal wind term and the effect of anisotropy on $\mathbfss{T}_{r\theta}$.
That neglect causes their meridional vorticity balance to favour weaker relative shears for slowly-rotating systems.

The baroclinicity $\lambda$ is in agreement with (albeit uncertain) measurements of the solar pole-equator temperature difference~\citep{Teplitskaya2015}.
Unfortunately while there have been measurements of the meridional circulation in the Sun~\citep{2013ApJ...774L..29Z,2015ApJ...813..114R,2020ApJ...890...32S}, there are still significant disagreements between the different inversion techniques which make a direct comparison to our theory challenging.

In the rapidly rotating limit our scaling for the differential rotation is in good agreement with three-dimensional MHD simulations of rapidly rotating solar-type stars~\citep{2017ApJ...836..192B} and stars with convecting cores~\citep{2016csss.confE.152A}.
In particular, we obtain differential rotation which increases sub-linearly with $\Omega$~\citep{2015ApJ...806...10M}.
We also find that the convection becomes increasingly magnetically dominated towards lower Rossby number.
This is typically seen in these simulations~\citep{2017ApJ...836..192B} and arises because the Coriolis effect arrests convective motions but does not impede the growth of the magnetic field.

We obtain a similar result in the hydrodynamic regime, where we again find qualitative agreement between our predicted shear and what is found in three-dimensional hydrodynamic simulations of rapidly rotating solar-type stars~\citep{2008ApJ...689.1354B,ASNA:ASNA201111624,2011A&A...531A.162K}, though our results disagree with at least some two-dimensional simulations~\citep{doi:10.1063/1.868716}.

In Fig.~\ref{fig:summary} we summarize our results for the differential rotation visually alongside data for the Sun and Jupiter.
These generally agree with the trends we find.
Interestingly, Jupiter seems to follow our hydrodynamic scaling law in the upper regions (left, red).
Further down (right, purple) Juno provides only upper bounds.
The transition between these regimes occurs at a depth quite similar to that at which ionization occurs.
The upper bounds are consistent with both scaling laws but~\citet{2018Natur.555..227G} suggest that the transition between the region with measurements and that with upper bounds is relatively sharp.
This could mean that at the depth where the atmosphere ionizes it also transitions from hydrodynamic to MHD scaling, resulting in a steep drop in shear.

In the slow-rotation limit we did not find any order of magnitude preference for different shear directions.
In that limit the angular momentum balance is chiefly between the turbulent viscosity, the thermal wind term and the $\Lambda$-effect, with all other terms scaling more slowly with angular velocity.

By contrast in the rapid-rotation limit the shear is preferentially in the cylindrical radial direction, and that preference is enforced by a factor of $\Omega/|N|$.
Specifically, in the azimuthal vorticity equation inertial or advective terms come to balance the thermal wind term, but the former preferentially couple to shear in the $\boldsymbol{e}_z$ direction and hence the system tends to preferentially shear orthogonal to this.
By contrast in the meridional vorticity equation both directions of the shear appear and balance against turbulent stresses, which have no such direction dependence.
As a result the shear perpendicular to the rotation axis is less constrained than that along it, tending to a Taylor-Proudman state~\citep{1897RSPTA.189..201H} of rotation on cylinders.
%%%---------- close: summary

%%%%%%%%%%% jump to conclusions
%%%---------- open: conclusions

\section{Conclusions}
\label{sec:conc}

We have made predictions for the scaling of the differential rotation, meridional circulation, magnetic field and baroclinic angle.
We are aware that three of our predicted scalings have been suggested previously.
These are that for the magnetic field energy~\citep{doi:10.1111/j.1365-246X.2006.03009.x}, that of the magnitude of differential rotation in the rapid hydrodynamic limit~\citep{2011Icar..211.1258S} and that of convection speed with rotation rate the hydrodynamic limit~\citep{1979GApFD..12..139S}.
The remaining scalings are new.

Our predictions suggest that a great many details of a convecting system are irrelevant to the question of the magnitude of its differential rotation, baroclinicity and meridional circulation.
All that matters to leading order is the bulk ratio $\Omega/|N|_0$, which is readily computed from observable parameters such as the rotation rate, surface temperature and mean density.
This allows a wide range of systems to be approximately characterised with minimal data and should help to support future studies of momentum and magnetic flux transport in convection zones.

Our finding that the shear is at most of order unity in slowly-rotating systems and is suppressed in rapid rotators places a strong bound on the amount of shear even a deep convection zone may support.
In particular, even for very deep regions, it is difficult to arrange for the angular velocity to vary across the zone by more than of order $|N|$.
This is because shear of that order requires that one boundary of the zone or the other is rotating rapidly, and so the achievable shear drops off quickly in that limit.

This bound is particularly important in red giants, for which the \brvs\ frequency corresponds to periods of order $100\,\rm{d}$ or more.
If the observed core-envelope shear in these systems is primarily in the convective envelope and not the interior radiative layer then the relative shear needs to be of order unity~\citep{0004-637X-808-1-35}, and this requires that the absolute shear be less than of order $|N|$.
Because the time-scale $|N|^{-1}$ is so long and the observed shear is so large, this favors scenarios which place significant shear in the radiative zones of red giants~\citep{2019MNRAS.485.3661F}.%%%---------- close: conclusions

\section*{Acknowledgements}

%%%%%%%%%%% jump to Snippets/acknowledgements
%%%---------- open: Snippets/acknowledgements
The Flatiron Institute is supported by the Simons Foundation. ASJ thanks the Gordon and Betty Moore Foundation (Grant GBMF7392) and the National Science Foundation (Grant No. NSF PHY-1748958) for supporting this work.
%%%---------- close: Snippets/acknowledgements
%%%%%%%%%%% jump to acknowledgements
%%%---------- open: acknowledgements
ASJ also acknowledges financial support from a UK Marshall Scholarship as well as from the IOA, ENS and CEBS to work at the IOA, ENS Paris and CEBS in Mumbai.
PL acknowledges travel support from the french PNPS (Programme National de Physique Stellaire) and from CEBS.
CAT thanks Churchill College for his fellowship.
SMC is grateful to the IOA for support and hospitality and thanks the Cambridge-Hamied exchange program for financial support.
ASJ, SMC and PL thank Bhooshan Paradkar for productive conversations related to this work.
The authors thank Jim Fuller, Chris Thompson, Douglas Gough and Steven Balbus for comments on this manuscript.
ASJ is grateful to Frank Timmes and Matteo Cantiello for suggestions regarding the presentation of this work.
%%%---------- close: acknowledgements

\section*{Data Availability}

The software used in this analysis is available at~\url{https://doi.org/10.5281/zenodo.3967763}.

\bibliographystyle{mnras}
\bibliography{refs}

\appendix

%%%%%%%%%%% jump to vorticity_appen
%%%---------- open: vorticity_appen
\section{Vorticity}
\label{appen:vort}

To pass from equations~\eqref{eq:vorticity} to equation~\eqref{eq:vorticity1} note that with equation~\eqref{eq:meridional_flow_breakout} we have
\begin{align}
	\boldsymbol{\omega}\cdot\nabla\boldsymbol{\varv}
	&= \boldsymbol{\omega}\cdot\nabla\boldsymbol{u} - \boldsymbol{e}_R \Omega \omega_\phi + \boldsymbol{e}_\phi \boldsymbol{\omega}\cdot\nabla (\Omega R),\\
	\boldsymbol{\varv}\cdot\nabla\boldsymbol{\omega}
	&= \boldsymbol{u}\cdot\nabla\boldsymbol{\omega} - \Omega \omega_\phi \boldsymbol{e}_R + \Omega \omega_R \boldsymbol{e}_\phi,\\
	\intertext{and}
	\boldsymbol{\omega}\nabla\cdot\boldsymbol{\varv}	
	&= \boldsymbol{\omega}\nabla\cdot\boldsymbol{u}.
\end{align}
Inserting these into equation~\eqref{eq:vorticity} recovers equation~\eqref{eq:vorticity1}.

%%%---------- close: vorticity_appen
%%%%%%%%%%% jump to entropy
%%%---------- open: entropy
\section{Entropy}
\label{appen:entropy}

The dimensionful entropy of an ideal gas per particle is given by the Sacker-Tetrode equation as
\begin{align}
	S = k_{\rm B} \left[\ln\left(\frac{\mu}{\rho}\left(\frac{4\pi \mu u}{3 h^2}\right)^{3/2}\right) + \frac{5}{2}\right],
\end{align}
where $u$ is the internal energy of the gas per particle, $\rho$ is its mass density, $h$ is Planck's constant, $k_{\rm B}$ is Boltzmann's constant and $\mu$ is the mean molecular weight of the particles~\citep{doi:10.1002/andp.19133450103,doi:10.1002/andp.19123430708}.
We non-dimensionalize this by letting
\begin{align}
	s \equiv \frac{S}{k_{\rm B}} = \ln\left(\frac{\mu}{\rho}\left(\frac{4\pi \mu u}{3 h^2}\right)^{3/2}\right) + \frac{5}{2}.
\end{align}
The internal energy of an ideal gas is proportional to its temperature, and hence to $P/\rho$, so
\begin{align}
	s \equiv \frac{S}{k_{\rm B}} = \ln\left(\frac{\mu}{\rho}\left(\frac{4\pi \mu P}{3 \rho h^2}\right)^{3/2}\right) + \mathrm{const.}
\end{align}
Treating $\mu$ as a constant, we may absorb all factors other than $\rho$ and $P$ into the additive constant.
We then set this to zero by appropriate choice of units for $P$ and $\rho$, leaving just
\begin{align}
	s = \frac{3}{2}\ln\frac{P}{\rho^{5/3}},
\end{align}
which for a monatomic ideal gas is equal to
\begin{align}
	s = \frac{1}{\gamma-1}\ln\frac{P}{\rho^\gamma},
\end{align}
which holds more generally for any ideal gas of constant $\gamma$.%%%---------- close: entropy
%%%%%%%%%%% jump to baroclinic
%%%---------- open: baroclinic
\section{Baroclinicity and Circulation}
\label{sec:baro}

\subsection{$\xi$ and $\lambda$}
\label{subsec:xilambda}

We begin with equations~\eqref{eq:xi},~\eqref{eq:lambda} and~\eqref{eq:s}, which may be combined to yield
\begin{align}
	\lambda &= \frac{|\nabla \ln P \times \frac{\gamma-1}{\gamma}\nabla s|}{|\nabla \ln P| |\nabla \ln \rho|}= \frac{(\gamma-1)|\nabla s|}{\gamma|\nabla \ln \rho|}\xi.
	\label{eq:lambda_inter}
\end{align}
We next define the \brvs\ frequency
\begin{align}
	N^2 \equiv -\frac{\gamma-1}{\gamma} \boldsymbol{g}\cdot\nabla s.
	\label{eq:brvs}
\end{align}
Expanding equation~\eqref{eq:brvs} we see that
\begin{align}
	N^2 &= -\frac{\gamma-1}{\gamma} \boldsymbol{g}\cdot\nabla s= -\frac{\gamma-1}{\gamma\rho}\nabla P \cdot \nabla s.
\end{align}
In the case of a convectively unstable entropy gradient this may be written as
\begin{align}
	N^2	= -\frac{(\gamma-1)P}{\gamma\rho}|\nabla \ln P| |\nabla s|\bar{\xi}.
	\label{eq:brvs_inter}
\end{align}
Rewriting equation~\eqref{eq:lambda_inter} in terms of $N$ using equation~\eqref{eq:brvs_inter} yields
\begin{align}
	\lambda &= \frac{-(\gamma-1)\rho N^2}{P|\nabla \ln \rho||\nabla \ln P|}\frac{\xi}{\bar{\xi}}.
	\label{eq:lambdaX}
\end{align}

When $\nabla P$ and $\nabla \rho$ are aligned, $|\nabla \ln \rho| \approx \gamma^{-1} |\nabla \ln P|$ because convection enforces a near-adiabatic relation.
When they are not the density gradient projected along the pressure gradient remains adiabatic because convective motions are primarily along the pressure gradient\footnote{This is because the pressure gradient is the source of the convective restoring force. See e.g. Fig. 16 of~\citet{jermyn}.}, so we instead obtain
\begin{align}
	|\nabla \ln \rho| &= \left|\frac{1}{\gamma} |\nabla \ln P| \left(\boldsymbol{e}_p + \frac{\lambda}{\bar{\lambda}}\boldsymbol{e}_q\right)\right|\\
	&=\frac{1}{\gamma \bar{\lambda}}|\nabla \ln P|
	\label{eq:rhop}
\end{align}
With equation~\eqref{eq:lambdaX} we obtain
\begin{align}
	\frac{\lambda}{\bar{\lambda}} = -\frac{\gamma\rho N^2}{p |\nabla \ln P|^2}\frac{\xi}{\bar{\xi}}.
\end{align}
With equation~\eqref{eq:magP} we find
\begin{align}
	\frac{\lambda}{\bar{\lambda}} &= -\frac{\gamma h^2 N^2 \rho}{P}\frac{\xi}{\bar{\xi}}.
\end{align}
Inserting equation~\eqref{eq:h} then yields
\begin{align}
	\frac{\lambda}{\bar{\lambda}} &= \gamma\frac{h N^2}{g}\frac{\xi}{\bar{\xi}},
	\label{eq:lambdaxi}
\end{align}
where
\begin{align}
	\label{eq:magP}
	h &\equiv |\nabla \ln P|^{-1}\\
	\label{eq:h}
	 &= \frac{P}{\rho g}
\end{align}
is the pressure scale height~\footnote{Equation~\eqref{eq:h} comes from hydrostatic equilibrium. In systems with circulation currents, stresses, and differential rotation there are deviations from this, but these are only large when the rotation is nearly Keplerian, and so we may ignore such corrections except in that regime.}.

A further simplification we shall often use is to let $\xi < \bar{\xi}$ and $\lambda < \bar{\lambda}$, which imply respectively that $\bar{\xi} \approx 1$ and $\bar{\lambda} \approx 1$.
This follows because were it not the case $\nabla P$ and $\nabla \rho$ would need to be nearly perpendicular, which can only be the case near breakup rotational velocities.
We address that case separately in section~\ref{sec:breakup}.

\subsection{Mass Conservation}
\label{subsec:masscons}

We expand the equation of mass conservation~\eqref{eq:masscons} as
\begin{align}
	0 &= \nabla \cdot (\rho \boldsymbol{u})\\
	&= \boldsymbol{u}\cdot\nabla\ln \rho + \nabla\cdot\boldsymbol{u}.
\end{align}
We may evaluate the first term as
\begin{align}
	\boldsymbol{u}\cdot\nabla\ln \rho &= (u_p \boldsymbol{e}_p + u_q \boldsymbol{e}_q)\cdot\nabla\ln\rho\\
	&= h(u_p \nabla\ln p + u_q \boldsymbol{e}_\phi \times \nabla \ln p)\cdot\nabla\ln\rho\\
	&= |\nabla \ln \rho|(\lambda u_q + u_p \bar{\lambda})\\
	&= (h\gamma)^{-1}\left(\frac{\lambda}{\bar{\lambda}} u_q + u_p\right)
	\label{eq:lnr}
\end{align}
The second may be expanded as
\begin{align}
	\nabla \cdot \boldsymbol{u} &= \nabla \cdot \left(\boldsymbol{e}_p u_p + \boldsymbol{e}_q u_q\right)\\
	&= \boldsymbol{e}_p \cdot \nabla u_p + \boldsymbol{e}_q \cdot \nabla u_q + u_p \nabla\cdot\boldsymbol{e}_p + u_q \nabla\cdot\boldsymbol{e}_q.
\end{align}
Putting these together we find
\begin{align}
	0 &= (h\gamma)^{-1}\left(\frac{\lambda}{\bar{\lambda}} u_q + u_p\right) + \boldsymbol{e}_p \cdot \nabla u_p + \boldsymbol{e}_q \cdot \nabla u_q + u_p \nabla\cdot\boldsymbol{e}_p + u_q \nabla\cdot\boldsymbol{e}_q.
	\label{eq:masscons2}
\end{align}

Because $\boldsymbol{e}_p$ is parallel to the gravitational field its variation is set by the overall scale and symmetry of the system, rather than local thermodynamic properties.
The unit vector $\boldsymbol{e}_q$ is then determined as a linear function of $\boldsymbol{e}_p$ and so varies over the same scales.
Thus, we expect that $\nabla\cdot\boldsymbol{e}_p \approx \nabla \cdot \boldsymbol{e}_q \approx 1/r$, where $r$ is the spherical radial scale and enters owing to the large-scale structure of the system.
Note that in the simple case where $\boldsymbol{e}_p = \boldsymbol{e}_r$ we have
\begin{align}
	\nabla\cdot\boldsymbol{e}_p &= \frac{2}{r}\\
	\nabla\cdot\boldsymbol{e}_q &= \frac{\cot\theta}{r}.
\end{align}

On the other hand $u_p$ and $u_q$ generally vary more rapidly than this because equation~\eqref{eq:heat2} shows that the velocity is related to the local condition of thermal equilibrium.
As a result we expect that the variation of $u_p$ and $u_q$ is due to the variation of local thermodynamic properties as well as global effects having to do with the scale and symmetry of the system.
Along the pressure gradient then
\begin{align}
	|\boldsymbol{e}_p \cdot \nabla \ln u_p| \approx |\boldsymbol{e}_p \cdot \nabla \ln u_q| \approx |\nabla \ln P| + r^{-1} = h^{-1} + r^{-1},
	\label{eq:ep}
\end{align}
where once more $r$ arises due to the large-scale structure of the system.
In general the pressure scale height is somewhat less than the radius, so
\begin{align}
	|\boldsymbol{e}_p \cdot \nabla \ln u_p| \approx h^{-1}.
	\label{eq:ep2}
\end{align}
Perpendicular to the pressure gradient only the density varies, so
\begin{align}
	|\boldsymbol{e}_q \cdot \nabla \ln u_p| &\approx |\boldsymbol{e}_q \cdot \nabla \ln u_q|\approx |\boldsymbol{e}_q\cdot\nabla \ln \rho| + r^{-1}\approx \frac{\lambda}{h} + \frac{1}{r},
	\label{eq:derivTheta}
\end{align}
where as before we used $\bar{\lambda} \approx 1$.
Inserting equations~\eqref{eq:ep},~\eqref{eq:ep2} and~\eqref{eq:derivTheta} into equation~\eqref{eq:masscons2} we find
\begin{align}
	\frac{1}{h}\left(|u_p| + \left(\lambda + \frac{h}{r}\right) |u_q|\right) \approx 0,
\end{align}
where we have neglected the signs of terms and just written down their magnitudes.
Note that we have used $\gamma \approx \bar{\lambda} \approx 1$, $h \ll r$ and $\lambda \ll 1$ but have not assumed anything about the relative magnitudes of $\lambda$ and $h/r$.
Because we assume minimal geometric tuning it must be that motion along $\boldsymbol{e}_p$ balances that along $\boldsymbol{e}_q$ in this equation, so
\begin{align}
	|u_p| \approx \left(\lambda + \frac{h}{r}\right) |u_q|.
	\label{eq:upq2}
\end{align}

\subsection{Thermal Equilibrium}
\label{subsec:thermo}

We expand $\boldsymbol{u}$ as
\begin{align}
	\boldsymbol{u} = \boldsymbol{e}_p u_p + \boldsymbol{e}_q u_q.
\end{align}
Inserting this and equation~\eqref{eq:grad_s} into equation~\eqref{eq:heat1} we obtain
\begin{align}
	u_p \bar{\xi} + u_q \xi = -\frac{\nabla\cdot\left(P \mathbfss{Q}\cdot\nabla s\right)}{P |\nabla s|}.
	\label{eq:heat2}
\end{align}
Taking $\bar{\xi} \approx 1$ and inserting equations~\eqref{eq:lambdaxi} and~\eqref{eq:upq} into equation~\eqref{eq:heat2} we find
\begin{align}
		|u_q| \xi \left[1 + \frac{h}{\xi r} + \frac{h N^2}{g}\right] + \frac{\left|\nabla\cdot\left(P \mathbfss{Q}\cdot\nabla s\right)\right|}{P |\nabla s|}\approx 0.	
\end{align}
We have assumed that the convection is efficient so $|N|^2 \ll g/h$ and we may drop that term to obtain
\begin{align}
		|u_q| \xi \left[1 + \frac{h}{\xi r}\right] + \frac{\left|\nabla\cdot\left(P \mathbfss{Q}\cdot\nabla s\right)\right|}{P |\nabla s|} \approx 0.	
\end{align}
Examining equation~\eqref{eq:upq} and using $h < r$ and $\xi < 1$ we see that up to factors of order unity $u \approx u_q$, so
\begin{align}
	\left(\xi + \frac{h}{r}\right) u + \frac{\left|\nabla\cdot\left(P \mathbfss{Q}\cdot\nabla s\right)\right|}{P |\nabla s|} \approx 0.
	\label{eq:uu}
\end{align}

\subsection{Thermal Wind}
\label{subsec:tw}

To obtain equation~\eqref{eq:tw1} from equation~\eqref{eq:tw0} we first insert equation~\eqref{eq:lambda} to find
\begin{align}
	|\rho^{-2}(\nabla P\times\nabla \rho)_\phi| &= \frac{P}{\rho}|\nabla \ln P||\nabla \ln \rho| \lambda.
\end{align}
Equation~\eqref{eq:rhop} then gives
\begin{align}
	|\rho^{-2}(\nabla P\times\nabla \rho)_\phi| &= \frac{P}{\rho}|\nabla \ln P|^2 \frac{\lambda}{\gamma\bar{\lambda}}.
\end{align}
Recalling equations~\eqref{eq:magP} and~\eqref{eq:h} we find
\begin{align}
	|\rho^{-2}(\nabla P\times\nabla \rho)_\phi| &= \frac{g}{h}\frac{\lambda}{\gamma\bar{\lambda}}.
\end{align}
Inserting equation~\eqref{eq:lambdaxi} we obtain equation~\eqref{eq:tw1}.

\subsection{Slowly Rotating Perturbations ($\Omega \ll |N|$)}
\label{subsec:pert_slow}

Next we analyze the effect of slow rotation on equation~\eqref{eq:heat1}.
To do so we note that the diffusivity tensor is of the form~\citep{2005A&A...431..345R,2013MNRAS.431.2200L}
\begin{align}
	\mathbfss{Q} \approx
h^2 |N|^{-1}
\begin{pmatrix}
|N|^2 & \Omega^2 \\ 
\Omega^2 & |N|^2
\end{pmatrix},
\label{eq:QQ}
\end{align}
where we have neglected multiplicative factors of order unity, the first column and row reflect $\boldsymbol{e}_p$ and the second of each reflects $\boldsymbol{e}_q$.
There is a second-order contribution of the form $\Omega|R\nabla\Omega|/|N|^2$, analogous to that in $\mathbfss{T}$, but we shall argue that $|R\nabla\Omega|$ is no greater than $\Omega$ and so absorb that contribution into the $\Omega^2$ terms.
Similarly there is a contribution from the baroclinicity proportional to $\xi$, but we shall show that this is at most of order $\Omega^2/|N|^2$ and so may likewise absorb it into the $\Omega^2$ terms.

With equation~\eqref{eq:QQ} we find
\begin{align}
	\nabla\cdot(P\mathbfss{Q}\cdot\nabla s)&= \nabla\cdot\left[\frac{h^2}{|N|} P |\nabla s| \left(\boldsymbol{e}_p (|N|^2 \bar{\xi} + \Omega^2 \xi) + \boldsymbol{e}_q (\Omega^2 \bar{\xi} + |N|^2 \xi)\right)\right].
	\label{eq:slow0}
\end{align}
When $\Omega=0$ the system is spherically symmetric and this vanishes, so 
\begin{align}
	\nabla\cdot(P\mathbfss{Q}\cdot\nabla s)&= \nabla\cdot\left[\frac{h^2}{|N|} P |\nabla s| \boldsymbol{e}_p |N|^2\right] = 0.
\end{align}
Subtracting this from equation~\eqref{eq:slow0} we find
\begin{align}
	\nabla\cdot(P\mathbfss{Q}\cdot\nabla s)= \nabla\cdot &\left[\frac{h^2}{|N|} P |\nabla s| \left(\boldsymbol{e}_p (|N|^2 (\bar{\xi}-1) + \Omega^2 \xi)\right.\right.\nonumber\\
	&\left.\left. + \boldsymbol{e}_q (\Omega^2 \bar{\xi} + |N|^2 \xi)\right)\right].
\end{align}
We expand $\bar{\xi}\approx 1$ and $\bar{\xi}-1 \approx -\xi^2/2$ as above to find
\begin{align}
	\nabla\cdot(P\mathbfss{Q}\cdot\nabla s)= \nabla\cdot &\left[\frac{h^2}{|N|} P |\nabla s| \left(\boldsymbol{e}_p \left(-\frac{1}{2}\xi^2 |N|^2 + \Omega^2 \xi\right)\right.\right.\nonumber\\
	&\left.\left. + \boldsymbol{e}_q (\Omega^2+ |N|^2 \xi)\right)\right].
	\label{eq:xi_int}
\end{align}

Each term on the right-hand side of equation~\eqref{eq:xi_int} is of the form
\begin{align}
	\frac{h^2}{|N|} P |\nabla s|&\nabla\cdot\left(\boldsymbol{e}_i f\left(\xi, |N|, \Omega, h, |N|, P, |\nabla s|\right) \right) \nonumber\\
	 	&=\frac{h^2}{|N|} P |\nabla s|\left(\nabla\cdot\boldsymbol{e}_i +\boldsymbol{e}_i\cdot\nabla f\left(\xi, |N|, \Omega, h, |N|, P, |\nabla s|\right) \right),
	 	\label{eq:form}
\end{align}
where $f$ is a product of powers of its arguments and $i$ is one of $p$ or $q$.
As argued in the text following equation~\eqref{eq:masscons2}, the divergence of our unit vectors is of order $r^{-1}$.
The other term in equation~\eqref{eq:form} may be expanded as
\begin{align}
	\boldsymbol{e}_i\cdot\nabla f &= \boldsymbol{e}_i\cdot\left(\nabla \ln p \left.\frac{\partial f}{\partial \ln p}\right|_{\ln \rho}+\nabla \ln \rho \left.\frac{\partial f}{\partial \ln \rho}\right|_{\ln p}\right)\\
	&= \boldsymbol{e}_i\cdot\left(- \frac{\boldsymbol{e}_p}{h}\left.\frac{\partial f}{\partial \ln p}\right|_{\ln \rho}+\nabla \ln \rho \left.\frac{\partial f}{\partial \ln \rho}\right|_{\ln p}\right).
	\label{eq:form_f}
\end{align}
Next we evaluate
\begin{align}
	\nabla \ln \rho \approx \frac{\bar{\lambda} \boldsymbol{e}_p}{h}  + \frac{\boldsymbol{e}_q \lambda}{h} + \frac{\boldsymbol{e}_q}{r}.
\end{align}
The first two terms just come from our definition of $\lambda$ and $\bar{\lambda}$ in equations~\eqref{eq:lambda} and~\eqref{eq:bar_lambda} respectively, along with the fact that $|\nabla \ln \rho|$ is of order $h^{-1}$.
The final term arises because the quantities being differentiated are perturbations driven by rotation and so are sensitive to the spherical geometry.
In principle there is also a term of order $\boldsymbol{e}_p / r$ for the same reason, but we omit that because it is smaller than the main contribution along the pressure gradient.
With $\bar{\lambda} \approx 1$ we then find
\begin{align}
	\nabla \ln \rho \approx \frac{\boldsymbol{e}_p}{h}  + \frac{\boldsymbol{e}_q \lambda}{h} + \frac{\boldsymbol{e}_q}{r}.
\end{align}
Inserting this into equation~\eqref{eq:form_f} we find
\begin{align}
	\boldsymbol{e}_i\cdot\nabla f & = \frac{1}{h}\boldsymbol{e}_i\cdot\left(\boldsymbol{e}_p  + \boldsymbol{e}_q \left(\lambda + \frac{h}{r}\right)\right).
	\label{eq:form_f}
\end{align}
With this equation~\eqref{eq:xi_int} may be written as
\begin{align}
	\nabla\cdot(P\mathbfss{Q}\cdot\nabla s)= &\frac{h}{|N|} P |\nabla s|\left(-\frac{1}{2}\xi^2 |N|^2 + \Omega^2 \xi + \left(\lambda + \frac{h}{r}\right)\right.\nonumber\\
	&\left.\times \left(\Omega^2+ |N|^2 \xi\right)\right).
		\label{eq:intermediate0}
\end{align}
Inserting this into equation~\eqref{eq:u}, dropping factors of order unity and neglecting signs we obtain
\begin{align}
		\left(\xi + \frac{h}{r}\right)u + \frac{h}{|N|}\left(\xi + \lambda + \frac{h}{r}\right) \left(\Omega^2+ |N|^2 \xi\right) &\approx 0,
		\label{eq:intermediate1}
\end{align}
Because $\lambda \ll \xi$ in convection zones where $|\nabla s| \ll |\nabla \ln P|$ this reduces to
\begin{align}
		\left(\xi + \frac{h}{r}\right)u + \frac{h}{|N|}\left(\xi + \frac{h}{r}\right) \left(\Omega^2+ |N|^2 \xi\right) &\approx0,
\end{align}
Dividing through by $\xi + h/r$ we find
\begin{align}
		u + \frac{h}{|N|}\left(\Omega^2+ |N|^2 \xi\right) &\approx0,
\end{align}
which is our condition of thermal equilibrium.
Note that we could have also considered perturbations to $\mathbfss{Q}$ owing to the differential rotation.
Like those owing to the rotation these also enter at second order, and break the same symmetries, so we may modify our result to
\begin{align}
		u + \frac{h}{|N|}\left(\left(|R\nabla\Omega| + \Omega\right)^2+ |N|^2 \xi\right) &\approx0.
		\label{eq:u_slow_2}
\end{align}

\subsection{Rapidly Rotating Perturbations ($\Omega \gg |N|$)}
\label{subsec:pert_fast}

Next we analyze the effect of rapid rotation on equation~\eqref{eq:heat1}.
In addition to its contribution to the stress tensor, the baroclinicity $\xi$ enters into the vorticity balance both in relation to the scale of the meridional circulation and by means of the thermal wind term.
We therefore wish to understand the various terms in the heat equation which depend on $\xi$.

The meridional circulation is given by equation~\eqref{eq:u} as
\begin{align}
	\left(\xi + \frac{h}{r}\right) u + \frac{\left|\nabla\cdot\left(P \mathbfss{Q}\cdot\nabla s\right)\right|}{P |\nabla s|} \approx 0.
	\label{eq:u_hydro0}
\end{align}
Our aim is to evaluate the second term as a function of $u$, $\xi$, $\Omega$ and $|R\nabla\Omega|$.

Unlike the limit of slow rotation the divergence of the convective flux no longer vanishes by symmetry considerations, so when $\xi = u = |R\nabla\Omega| = 0$ we expect that
\begin{align}
\frac{\left|\nabla\cdot\left(P \mathbfss{Q}\cdot\nabla s\right)\right|}{P |\nabla s|} \approx \frac{\mathsf{Q}}{h}.
\end{align}
Without any symmetry to preclude this we expect the baroclinicity, meridional circulation and shear to perturb it at first order, so when these are non-zero
\begin{align}
\frac{\left|\nabla\cdot\left(P \mathbfss{Q}\cdot\nabla s\right)\right|}{P |\nabla s|} \approx \frac{\mathsf{Q}}{h}\left(1 + \xi + \frac{|R\nabla\Omega|}{|N|} + \frac{u}{h|N|}\right),
\label{eq:flux_div}
\end{align}
where, as usual, we have normalized the perturbations by the convective time-scale $|N|$.

We use mixing length theory and find the diffusivity to be
\begin{align}
	\mathsf{Q} \approx h \varv_{\rm c}.
	\label{eq:MHDQ}
\end{align}
Using equation~\eqref{eq:fastVC} we find
\begin{align}
	 \mathsf{Q} \approx h^2 |N| k
\end{align}
So equation~\eqref{eq:flux_div} becomes
\begin{align}
\frac{\left|\nabla\cdot\left(P \mathbfss{Q}\cdot\nabla s\right)\right|}{P |\nabla s|} \approx k h |N|\left(1 + \xi + \frac{|R\nabla\Omega|}{|N|} + \frac{u}{h|N|}\right).
\end{align}
Inserting this into equation~\eqref{eq:u_hydro0} we obtain
\begin{align}
	\left(\xi + \frac{h}{r} + k\right) \frac{u}{h|N|} + \left(1 + \xi + \frac{|R\nabla\Omega|}{\Omega}\right) k \approx 0.
\end{align}
We shall later determine that $\xi$ is of order unity in this limit, so the factor of $h/r$ may be dropped to yield
\begin{align}
	\left(\xi + k\right) \frac{u}{h|N|} + \left(1 + \xi + \frac{|R\nabla\Omega|}{\Omega}\right) k \approx 0.
	\label{eq:u_MHD11}
\end{align}

%%%---------- close: baroclinic
%%%%%%%%%%% jump to stress_terms
%%%---------- open: stress_terms

\section{Stress Terms}
\label{appen:stress}

\subsection{Derivatives}
\label{appen:curl}

In spherical coordinates we write
\begin{align}
	\mathbfss{T} =
	\begin{pmatrix}
		\mathsf{T}_{rr} & \mathsf{T}_{r\theta} & \mathsf{T}_{r\phi}\\
		\mathsf{T}_{\theta r} & \mathsf{T}_{\theta\theta} & \mathsf{T}_{\theta\phi}\\
		\mathsf{T}_{\phi r} & \mathsf{T}_{\phi\theta} & \mathsf{T}_{\phi\phi}		
	\end{pmatrix}.
\end{align}
Using Mathematica we expand the stress terms appearing in equation~\eqref{eq:vorticity} and find
\begin{align}
\label{eq:dTr}
	\nabla\times\left(\frac{1}{\rho}\nabla\cdot\mathbfss{T}\right)_r
	&= \frac{1}{r^2 \rho^2}\left[\rho \partial_\theta \mathsf{T}_{\phi r}+\mathsf{T}_{\phi r} \left(\cot \theta \rho-\partial_\theta\rho\right)\right.\nonumber\\
	&\left.+\cot \theta \rho \partial_\theta\mathsf{T}_{\phi\theta}-\mathsf{T}_{\phi\theta} \left(\cot \theta \partial_\theta \rho+\rho\right)\right.\nonumber\\
	&\left.\mathcolorbox{green}{-r \partial_\theta \rho \partial_r \mathsf{T}_{ r\phi}}+2 \rho \partial_\theta\mathsf{T}_{ r\phi}+\mathcolorbox{green}{r \rho \partial_r\partial_\theta\mathsf{T}_{ r\phi}}\right.\nonumber\\
	&\left.+\mathcolorbox{green}{r \cot \theta \rho \partial_r \mathsf{T}_{ r\phi}}-2 \partial_\theta\rho \mathsf{T}_{ r\phi}+2 \cot \theta \rho \mathsf{T}_{ r\phi}\right.\nonumber\\
	&\left.-\partial_\theta \rho \partial_\theta\mathsf{T}_{\theta\phi }+\rho \partial_\theta^2\mathsf{T}_{\theta\phi }+2 \cot \theta \rho \partial_\theta\mathsf{T}_{\theta\phi }\right.\nonumber\\
	&\left.-\cot \theta \partial_\theta \rho \mathsf{T}_{\theta\phi }-\rho \mathsf{T}_{ \theta\phi}\right],\\
\label{eq:dTt}
	\nabla\times\left(\frac{1}{\rho}\nabla\cdot\mathbfss{T}\right)_\theta
	&=	\frac{1}{r \rho^2}\left[\partial_r \rho \left(\mathsf{T}_{\phi r}+\cot \theta (\mathsf{T}_{\phi\theta}+\mathsf{T}_{ \theta\phi})\right.\right.\nonumber\\
	&\left.\left.+r \partial_r\mathsf{T}_{r\phi }+2 \mathsf{T}_{ r\phi}+\partial_\theta\mathsf{T}_{\theta\phi }\right)\right.\nonumber\\
	&\left.-\rho r \left(\partial_r\mathsf{T}_{\phi r}+\cot \theta \left(\partial_r\mathsf{T}_{\phi\theta}+\partial_r\mathsf{T}_{ \theta\phi}\right)\right.\right.\nonumber\\
	&\left.\left.+3 \partial_r\mathsf{T}_{r\phi }+\mathcolorbox{green}{r \partial_r^2\mathsf{T}_{r\phi }}+\partial_r\partial_\theta\mathsf{T}_{\theta\phi }\right)\right]\\
\mathrm{and}\nonumber\\
\label{eq:dTp}
	\nabla\times\left(\frac{1}{\rho}\nabla\cdot\mathbfss{T}\right)_\phi
	&= \frac{1}{r^2 \rho^2}\left[\rho \left(-2 \partial_\theta\mathsf{T}_{rr}-\partial_\theta^2\mathsf{T}_{\theta r}\right.\right.\nonumber\\
	&\left.\left.-\cot \theta \partial_\theta\mathsf{T}_{\theta r}+\partial_\theta\mathsf{T}_{\theta \theta}+\partial_\theta\mathsf{T}_{\phi \phi}\right.\right.\nonumber\\
	&\left.+r \left(-\partial_r\partial_\theta\mathsf{T}_{rr}+\partial_r\mathsf{T}_{\theta r}+3 \partial_r\mathsf{T}_{ r \theta}+\mathcolorbox{green}{r \partial_r^2\mathsf{T}_{ r\theta}}\right.\right.\nonumber\\
	&\left.\left.+\partial_r\partial_\theta\mathsf{T}_{\theta \theta}+\cot \theta \left(\partial_r\mathsf{T}_{\theta \theta}-\partial_r\mathsf{T}_{\phi \phi}\right)\right)\right)\nonumber\\
	&+\partial_\theta \rho \left(r \partial_r\mathsf{T}_{rr}+2 \mathsf{T}_{rr}+\partial_\theta\mathsf{T}_{\theta r}-\mathsf{T}_{\theta \theta}-\mathsf{T}_{\phi \phi}\right)\nonumber\\
	&\left.+\mathsf{T}_{\theta r} \left(-r \partial_r \rho+\cot \theta \partial_\theta \rho+\csc ^2(\theta ) \rho\right)\right.\nonumber\\
	&\left.-r \partial_r \rho \left(r \partial_r\mathsf{T}_{r\theta }+2 \mathsf{T}_{ r\theta}+\partial_\theta\mathsf{T}_{\theta \theta}\right.\right.\nonumber\\
	&\left.\left.+\cot \theta (\mathsf{T}_{\theta \theta}-\mathsf{T}_{\phi \phi})\right)\right].
\end{align}
Inspection of these equations reveals that only the off-diagonal components of $\mathsf{T}$ contribute to the $r$ and $\theta$ components of the vorticity equation, while all but the $r\phi$, $\theta r$, $\phi r$, and $r \theta$ components of $\mathsf{T}$ contribute to the $\phi$ component of the vorticity equation.
The terms highlighted in green are the ones we find to be dominant in the slowly-rotating limit.

\subsection{Slow Scaling ($\Omega \ll |N|$)}
\label{subsec:slow}

When there is rotation, shear or baroclinicity, spherical symmetry is broken.
The stress tensor is then perturbed away from the symmetric form
\begin{align}
	\mathbfss{T} &\approx 
	\begin{pmatrix}
\mathsf{T}_{rr} & 0 & 0 \\ 
0 & \mathsf{T}_{\theta\theta} & 0 \\
0 & 0 & \mathsf{T}_{\phi\phi} 
\end{pmatrix}.
\end{align}
It is this perturbation which gives the contribution of the stress to equations~\eqref{eq:vortm0} and~\eqref{eq:vortp0}.
However these effects break the symmetry in different ways and so enter as perturbations at different orders.
In this section we shall evaluate the perturbations to $\nabla\times(\rho^{-1}\nabla\cdot\mathbfss{T})$ owing to rotation in the limit where $\Omega \ll |N|$.
We begin with perturbations owing to rotation directly, then consider those from shear in the $|R\nabla\Omega| \ll |N|$ limit, those from the meridional flow in the $|\nabla u| \ll |N|$ limit, and finally those from baroclinicity in the $\xi \ll 1$ limit.

\subsubsection{Rotation}
To leading order, the perturbation owing to rotation is of the form~\citep{1994ApJ...425..303C,2013IAUS..294..399K,jermyn}
\begin{align}
	\delta\mathbfss{T} & \approx
	\mathsf{T}\begin{pmatrix}
\frac{\Omega^2}{|N|^2} & \frac{\Omega^2}{|N|^2} & \frac{\Omega}{|N|} \\ 
\frac{\Omega^2}{|N|^2} & \frac{\Omega^2}{|N|^2} & \frac{\Omega}{|N|} \\
\frac{\Omega}{|N|} & \frac{\Omega}{|N|} & \frac{\Omega^2}{|N|^2}
\end{pmatrix},
\label{eq:TpertSlow}
\end{align}
where we have neglected dimensionless factors of order unity which multiply the various factors of $\Omega/|N|$.
The first order perturbations arise in the $r\phi$, $\phi r$, $\phi \theta$ and $\theta\phi$ components because the Coriolis effect couples motion along other directions to motion along $\boldsymbol{e}_\phi$, while the perturbations to the remaining components are second order because a second application of the Coriolis effect is required to couple motion along two directions neither of which is $\boldsymbol{e}_\phi$.
Finally, the perturbations on the diagonal are second order, both because of the centrifugal effect and because it takes two applications of the Coriolis effect to couple motion in a given direction to itself.

In appendix~\ref{appen:curl} we expanded the curl of the divergence of the turbulent stress.
Inspection of equations~\eqref{eq:dTr},~\eqref{eq:dTt} and~\eqref{eq:dTp} reveals that only the off-diagonal components of $\mathbfss{T}$ contribute to the $r$ and $\theta$ components of the vorticity equation, while all but the $r\phi$, $\theta r$, $\phi r$, and $r \theta$ components of $\mathbfss{T}$ contribute to the $\phi$ component of the vorticity equation.
As before we take radial gradients to produce factors of $h^{-1}$ and latitudinal gradients to produce factors of $r^{-1} + h^{-1} \lambda/\bar{\lambda} \approx r^{-1} + \lambda h^{-1}$.
Because $\lambda \ll 1$ and $h \ll r$, terms with fewer latitudinal derivatives are dominant.

In equation~\eqref{eq:dTr} there are several terms involving one radial derivative which are perturbed at first order, so these are the most important.
In equation~\eqref{eq:dTt} there is a term involving two derivatives which is perturbed at first order, so that is the dominant term.
In equation~\eqref{eq:dTp} there is one term which is perturbed at second (leading) order and which has only radial derivatives, so that is the important term.
These terms, which actually dominate in all symmetry-constrained circumstances we consider, are highlighted in green in equations~\eqref{eq:dTr},~\eqref{eq:dTt} and~\eqref{eq:dTp}.
With that, we obtain
\begin{align}
\label{eq:TrO}
	\nabla\times\left(\frac{1}{\rho}\nabla\cdot\mathbfss{T}\right)_r  &\approx \frac{1}{\rho h^2}\left(\lambda + \frac{h}{r}\right)\mathsf{T}\frac{\Omega}{|N|},\\
\label{eq:TtO}
	\nabla\times\left(\frac{1}{\rho}\nabla\cdot\mathbfss{T}\right)_\theta &\approx \frac{1}{\rho h^2}\mathsf{T}\frac{\Omega}{|N|}\\
\intertext{and}
\label{eq:TpO}
	\nabla\times\left(\frac{1}{\rho}\nabla\cdot\mathbfss{T}\right)_\phi &\approx \frac{1}{\rho h^2}\mathsf{T}\frac{\Omega^2}{|N|^2}.
\end{align}

\subsubsection{Differential Rotation}
Differential rotation likewise breaks spherical symmetry.
The $r\phi$, $\phi r$, $\theta \phi$ and $\phi\theta$ components of $\mathbfss{T}$ break this symmetry at first order because they directly couple to the shear, while the remaining terms break the symmetry at second order both by coupling to the shear twice and by coupling once to the shear and once to the rotation itself.
To see this note that the differential rotation acts in the plane of $\boldsymbol{e}_\phi$ and the shear direction, so if motions along $\boldsymbol{e}_r$, $\boldsymbol{e}_\theta$ and $\boldsymbol{e}_\phi$ are not correlated initially then motions between $\boldsymbol{e}_r$ and $\boldsymbol{e}_\theta$ cannot be coupled at first order by the differential rotation.
Thus either the Coriolis effect is needed to couple these components or else a higher order perturbation is needed.
For the same reason the differential rotation cannot perturb the diagonal components of $\mathbfss{T}$ to first order, and two applications are needed to turn motion along, say, $\boldsymbol{e}_r$ into motion along another axis and back into motion along $\boldsymbol{e}_r$.
Another way to understand this is to note that the mapping $\varv_\phi \rightarrow -\varv_\phi$ also maps $\Omega \rightarrow -\Omega$ and $\nabla\Omega \rightarrow -\nabla\Omega$, which may be undone by then letting $\phi \rightarrow -\phi$.
This spatial transformation negates the components of $\mathbfss{T}$ which involve the direction $\boldsymbol{e}_\phi$ an odd number of times but not those involving it an even number of times\footnote{One might ask why this transformation is not also undone by letting $z \rightarrow -z$. The reason is that while $\boldsymbol{\varv}$ is a vector, $\boldsymbol{\Omega}$ is a pseudovector generated by a cross-product with $\boldsymbol{e}_\phi$, so its component along the $\boldsymbol{e}_z$ axis is left invariant upon reflection about that axis.}, and so the latter must be even functions of $\Omega$.
It follows that they must be at least quadratic in $\Omega$ and hence only terms of the form $\Omega|R\nabla\Omega|$ and $|R\nabla\Omega|^2$ are allowed at leading order.
This is in agreement with the model of~\citet{1994AN....315..157K}.
Based on these symmetry arguments we write the leading order term in the differential rotation
\begin{align}
\label{eq:TrD}
	\nabla\times\left(\frac{1}{\rho}\nabla\cdot\mathbfss{T}\right)_r &\approx \frac{1}{\rho h^2}\left(\lambda + \frac{h}{r}\right)\mathsf{T}\frac{|R\nabla\Omega|}{|N|},\\
\label{eq:TtD}
	\nabla\times\left(\frac{1}{\rho}\nabla\cdot\mathbfss{T}\right)_\theta &\approx \frac{1}{\rho h^2}\mathsf{T}\frac{|R\nabla\Omega|}{|N|}\\
\intertext{and}
\label{eq:TpD}
	\nabla\times\left(\frac{1}{\rho}\nabla\cdot\mathbfss{T}\right)_\phi &\approx \frac{1}{\rho h^2}\mathsf{T}\frac{\Omega|R\nabla\Omega|}{|N|^2}.
\end{align}

\subsubsection{Meridional Circulation}
A similar argument produces the couplings to the meridional circulation.
Mapping $\phi \rightarrow -\phi$ leaves both the meridional circulation and its shear unchanged, yet it negates all components of the stress along $\boldsymbol{e}_\phi$.
All terms in equations~\eqref{eq:dTr} and~\eqref{eq:dTt} therefore vanish unless $\phi \rightarrow -\phi$ is a broken symmetry, which requires that one of $\Omega$ or $|R\nabla\Omega|$ be non-zero.
It follows that the meridional circulation perturbs the meridional vorticity equation via the stress following
\begin{align}
\label{eq:TrU}
	\nabla\times\left(\frac{1}{\rho}\nabla\cdot\mathbfss{T}\right)_r &\approx \frac{1}{\rho h^2}\left(\lambda + \frac{h}{r}\right)\mathsf{T}\frac{|\nabla\boldsymbol{u}|}{|N|}\left(\frac{\Omega}{|N|} + \frac{|R\nabla\Omega|}{|N|}\right)\\
\intertext{and}
\label{eq:TtU}
	\nabla\times\left(\frac{1}{\rho}\nabla\cdot\mathbfss{T}\right)_\theta &\approx \frac{1}{\rho h^2}\mathsf{T}\frac{|\nabla\boldsymbol{u}|}{|N|}\left(\frac{\Omega}{|N|} + \frac{|R\nabla\Omega|}{|N|}\right)
\end{align}
On the other hand, the same mapping of $\phi \rightarrow -\phi$ does not negate every term in equation~\eqref{eq:TpO}, for example $\partial_r^2 \mathsf{T}_{\theta r}$ is left unchanged.
Those terms may therefore be non-zero.
Under the mapping $\boldsymbol{e}_\theta \rightarrow -\boldsymbol{e}_\theta$ these are negated, just as the relevant component of the meridional flow is negated, so they can couple to the meridional circulation at first order.
It follows that the meridional circulation can appear in this stress contribution at first order without any intermediating effects, so
\begin{align}
\label{eq:TpU}
	\nabla\times\left(\frac{1}{\rho}\nabla\cdot\mathbfss{T}\right)_\phi &\approx \frac{1}{\rho h^2}\mathsf{T}\frac{|\nabla\boldsymbol{u}|}{|N|}.
\end{align}
That is, $\boldsymbol{u}$ only couples to the stress via the perturbed terms in the meridional vorticity equation whereas it couples directly in the azimuthal component.

Note that the couplings in equations~\eqref{eq:TrO} and ~\eqref{eq:TtO} are the leading order contributions to the $\Lambda$-effect, while those in equations~(\ref{eq:TrD})~to~(\ref{eq:TpU}) produce an effective turbulent viscosity with magnitude $|N|^{-1}\mathsf{T}$ \citep{1989drsc.book.....R, 2013IAUS..294..399K}.
These expansions are consistent with standard closure models such as those used by \citet{2012ISRAA2012E...2G} and \citet{2013MNRAS.431.2200L}, as well as with simulations of slowly rotating convection~\citep{2011A&A...531A.162K}.

\subsubsection{Baroclinicity}
There is one further effect which may contribute to the stress at leading order, namely baroclinicity~\citep{1989drsc.book.....R,jermyn}.
This effect lies inside the meridional plane and hence is invariant with respect to the mapping $\phi \rightarrow -\phi$.
It follows that it only contributes at leading order to components of $\mathbfss{T}$ which incorporate the direction $\boldsymbol{e}_\phi$ an even number of times.
This is the same case we dealt with for the coupling to the meridional circulation.
As a result
\begin{align}
\label{eq:TpB}
	\nabla\times\left(\frac{1}{\rho}\nabla\cdot\mathbfss{T}\right)_\phi &\approx \frac{1}{\rho h^2}\mathsf{T}\xi.
\end{align}
In order to couple $\xi$ into the remaining components of the vorticity equation we need other effects to break this symmetry.
Hence
\begin{align}
\label{eq:TrB}
	\nabla\times\left(\frac{1}{\rho}\nabla\cdot\mathbfss{T}\right)_r &\approx \frac{1}{\rho h^2}\left(\lambda + \frac{h}{r}\right)\mathsf{T}\xi\left(\frac{\Omega}{|N|} + \frac{|R\nabla\Omega|}{|N|}\right)\\
\intertext{and}
\label{eq:TtB}
	\nabla\times\left(\frac{1}{\rho}\nabla\cdot\mathbfss{T}\right)_\theta &\approx \frac{1}{\rho h^2}\mathsf{T}\xi\left(\frac{\Omega}{|N|} + \frac{|R\nabla\Omega|}{|N|}\right).
\end{align}

\subsubsection{Overall Perturbation}
Putting it all together with equation~\eqref{eq:Tslow} we find
\begin{align}
	\nabla\times\left(\frac{1}{\rho}\nabla\cdot\mathbfss{T}\right)_r &\approx \left(\lambda + \frac{h}{r}\right)|N|\left(\Omega+|R\nabla\Omega|\right)\left(1 + \xi + \frac{u}{h|N|}\right),\\
	\nabla\times\left(\frac{1}{\rho}\nabla\cdot\mathbfss{T}\right)_\theta &\approx |N|\left(\Omega+|R\nabla\Omega|\right)\left(1 + \xi + \frac{u}{h|N|}\right)\\
\intertext{and}
	\nabla\times\left(\frac{1}{\rho}\nabla\cdot\mathbfss{T}\right)_\phi &\approx |N|^2 \left(\xi + \frac{u}{h} + \frac{\Omega^2}{|N|^2} + \frac{\Omega|R\nabla\Omega|}{|N|^2}\right),
\end{align}
where we have used the fact from section~\ref{sec:merid} that $|\nabla\boldsymbol{u}| \approx u/h$ and have added additional terms at higher order to permit a more compact representation.
These are not necessarily present, though there is no symmetry which prohibits them.

\subsection{Rapid Scaling ($\Omega \gg |N|$)}
\label{subsec:rapid}

When $|R\nabla\Omega| \ll \varv_{\mathrm{c}}/h$ the turbulent stress is dominated by convective motions.
In the opposite limit it is dominated by the shear.
Taking $|R\nabla\Omega|$ to be the frequency scale of the shear forcing and $h$ to be its characteristic length, we write the shear turbulent velocity as
\begin{align}
	\varv_{\rm s} \approx h |R\nabla\Omega| k,
\end{align}
where $k \equiv \left(\frac{\Omega}{|R\nabla\Omega|}\right)$ analogously to the scaling of convection forced at $|N|$ in a rapidly rotating system~(equation~\ref{eq:fastVC}).
The extra factor of $k$ again comes from the Coriolis effect stabilizing motion perpendicular to the rotation axis.
In section~\ref{sec:stiff} we argued that $|R\nabla\Omega| \la |N|$.
Because of this,
\begin{align}
	\varv_{\rm s} \la |N|,
\end{align}
so the shear never dominates the stress.
We therefore have the usual convective result~\citep{gough78}
\begin{align}
	\mathsf{T} \approx \rho \varv_{\mathrm{c}}^2.
\end{align}

In the rapidly rotating regime the stress is not symmetry protected, because the rotation is rapid and the off-diagonal components of $\mathbfss{T}$ are of the same order as the diagonal~\citep{2013IAUS..294..399K,jermyn}.
Hence, in the absence of shear and baroclinicity, we write
\begin{align}
\label{eq:FTor}
	\nabla\times\left(\frac{1}{\rho}\nabla\cdot\mathbfss{T}\right)_r &\approx \left(\frac{h}{r} + \lambda\right)|N|^2 k^2,\\
\label{eq:FTot}
	\nabla\times\left(\frac{1}{\rho}\nabla\cdot\mathbfss{T}\right)_\theta &\approx |N|^2 k^2\\
\intertext{and}
\label{eq:FTop}
	\nabla\times\left(\frac{1}{\rho}\nabla\cdot\mathbfss{T}\right)_\phi &\approx |N|^2 k^2,
\end{align}
where we have followed the prescription in section~\ref{sec:slow} to evaluate the derivatives.

Similarly, the convective turbulence couples viscously to the differential rotation and any meridional shear.
Because there is no symmetry protection, we assume that this produces a coupling at first order.
The relevant diffusivity is just the stress divided by its characteristic time-scale, which is $|N|$ for convectively-dominated turbulence.
So,
\begin{align}
\label{eq:FTwr}
	\nabla\times\left(\frac{1}{\rho}\nabla\cdot\mathbfss{T}\right)_r &\approx \left(\frac{h}{r} + \lambda\right)|N| k^2 |R\nabla\Omega|,\\
\label{eq:FTwt}
	\nabla\times\left(\frac{1}{\rho}\nabla\cdot\mathbfss{T}\right)_\theta &\approx |N| k^2 |R\nabla\Omega|\\
\intertext{and}
\label{eq:FTwp}
	\nabla\times\left(\frac{1}{\rho}\nabla\cdot\mathbfss{T}\right)_\phi &\approx |N| k^2 |R\nabla\Omega|.
\end{align}
There is likewise a first-order viscosity-like coupling to the meridional circulation of the form 
\begin{align}
\label{eq:viscFastR}
	\nabla\times\left(\frac{1}{\rho}\nabla\cdot\mathbfss{T}\right)_r &\approx \left(\frac{h}{r} + \lambda\right)|N| k^2 |\nabla \boldsymbol{u}|,\\
\label{eq:viscFastT}
	\nabla\times\left(\frac{1}{\rho}\nabla\cdot\mathbfss{T}\right)_\theta &\approx |N| k^2 |\nabla \boldsymbol{u}|\\
\intertext{and}
\label{eq:viscFastP}
	\nabla\times\left(\frac{1}{\rho}\nabla\cdot\mathbfss{T}\right)_\phi &\approx |N| k^2 |\nabla \boldsymbol{u}|,
\end{align}
where $|\nabla \boldsymbol{u}|$ is the magnitude of the tensor formed of derivatives of the velocity components.

Finally, we must consider the contribution of baroclinicity to the stress.
Its presence breaks no symmetries in this limit, and over its possible range from $-1$ to $1$ it completely changes the character of the convection, so we approximate its effect as being linear and of order unity.
Combining this with equations~\eqref{eq:FTor},~\eqref{eq:FTwr} and~\eqref{eq:viscFastR} and letting $|\nabla \boldsymbol{u}| \approx u/h$ we find
\begin{align}
	\label{eq:hr1}
		\nabla\times\left(\frac{1}{\rho}\nabla\cdot\mathbfss{T}\right)_r &\approx k^2 |N| \left(|N| + \xi |N| + \frac{u}{h}+|R\nabla\Omega|\right) \left(\frac{h}{r} + \lambda\right),
\end{align}
Likewise, equations~\eqref{eq:FTot},~\eqref{eq:FTwt} and~\eqref{eq:viscFastT} give
\begin{align}
	\label{eq:hr2}
		\nabla\times\left(\frac{1}{\rho}\nabla\cdot\mathbfss{T}\right)_\theta &\approx k^2 |N| \left(|N| + \xi |N| + \frac{u}{h}+|R\nabla\Omega|\right),
\end{align}
and finally equations~\eqref{eq:FTop},~\eqref{eq:FTwp} and~\eqref{eq:viscFastP} produce
\begin{align}
	\label{eq:hr3}
		\nabla\times\left(\frac{1}{\rho}\nabla\cdot\mathbfss{T}\right)_\phi &\approx k^2 |N| \left(|N| + \xi |N| + \frac{u}{h}+|R\nabla\Omega|\right).
\end{align}

\section{Advective Terms}
\label{appen:tab2}

In this appendix we evaluate the terms in the vorticity equations~\eqref{eq:vortm1} and~\eqref{eq:vortp1} which capture the kinematic effects associated with the rotation and circulation

\subsection{Meridional Equation~\eqref{eq:vortm1}}

We begin with the term $\boldsymbol{\omega}_m\cdot\nabla_m\boldsymbol{u}$.
We may evaluate $\boldsymbol{\omega}_m$ by expanding equation~\eqref{eq:vort} with equation~\eqref{eq:meridional_flow_breakout} and projecting into the meridional plane to obtain
\begin{align}
	\boldsymbol{\omega}_m &= \boldsymbol{e}_r \left(2\cos\theta \Omega + \sin\theta \frac{\partial \Omega}{\partial \theta}\right) - \boldsymbol{e}_\theta \sin\theta \left(2\Omega + r\frac{\partial \Omega}{\partial r}\right),
	\label{eq:vortexp}
\end{align}
which does not depend on $\boldsymbol{u}$ because the vorticity is a curl and $\boldsymbol{u}$ is itself meridional.
Next, we write the meridional circulation $\boldsymbol{u}$ in the basis formed by the pressure gradient and the perpendicular unit vector in the meridional plane we obtain
\begin{align}
	\boldsymbol{u} = u_p \boldsymbol{e}_p + u_q \boldsymbol{e}_q.
\end{align}
When the system is slowly rotating, $\boldsymbol{e}_p \approx \boldsymbol{e}_r$ and $\boldsymbol{e}_q \approx \boldsymbol{e}_\theta$, with corrections to both of order $\lambda$.
Even in the limit of rapid rotation $\lambda \approx h |N|^2 / g \ll 1$, so we may neglect such corrections and write
\begin{align}
	\boldsymbol{\omega}_m\cdot\nabla\boldsymbol{u}
	&\approx \boldsymbol{e}_r\left[\partial_r u_r \left(2\Omega\cos\theta + \partial_\theta\Omega \sin\theta\right)\right.\nonumber\\
	&\left. + \sin\theta \left(u_\theta-\partial_\theta u_r\right)\left(2\Omega + r \partial_r \Omega\right)\right]\nonumber\\
	&+ \boldsymbol{e}_\theta\left[\partial_r u_\theta \left(2\Omega\cos\theta + \partial_\theta\Omega \sin\theta\right)\right.\nonumber\\
	&\left. - \sin\theta \left(u_r+\partial_\theta u_\theta\right)\left(2\Omega + r \partial_r \Omega\right)\right].
\end{align}
Making the approximation that radial derivatives of $\boldsymbol{u}$ produce factors of $h^{-1}$ while latitudinal ones produce factors of $r^{-1}$, taking $h \ll r$ and ignoring factors of order unity we find
\begin{align}
	\left|\boldsymbol{\omega}_m\cdot\nabla\boldsymbol{u}\right| &\approx \frac{1}{h} \left[\Omega u_r + \Omega u_\theta + u_r |R\nabla\Omega| + u_\theta |R\nabla\Omega|\right],
\end{align}
where we have replaced $r\partial_r \Omega$ by $|R\nabla\Omega|$.
Equation~\eqref{eq:upq} then tells us that $u \approx u_\theta \ll u_r$ so
\begin{align}
	\left|\boldsymbol{\omega}_m\cdot\nabla\boldsymbol{u}\right| &\approx \frac{u}{h} \left[\Omega + |R\nabla\Omega|\right].
	\label{eq:omu}
\end{align}
The relative corrections to this are at least of order $h/r$, $\lambda$ and $\Omega/|N|$.
Because the non-rotating system is spherically symmetric $\lambda$ must be at least of order $\Omega/|N|$ too. Hence this expansion is accurate to leading order in both $h/r$ and $\Omega/|N|$.

We next turn to the term $\boldsymbol{\omega}_m\boldsymbol{u}\cdot\nabla\ln\rho$, the magnitude of which is
\begin{align}
	|\boldsymbol{\omega}_m\boldsymbol{u}\cdot\nabla\ln\rho| &= |\boldsymbol{\omega}_m| |\boldsymbol{u}\cdot\nabla\ln\rho|.
\end{align}
The first term we may find using equation~\eqref{eq:vortexp} to be\
\begin{align}
	|\boldsymbol{\omega}_m| &\approx |R\nabla\Omega| + \Omega.
\end{align}
The remaining term we have already computed in equation~\eqref{eq:lnr} and found to be
\begin{align}
	|\boldsymbol{u}\cdot\nabla\ln\rho| \approx \frac{1}{h\gamma}\left(u_p + u_q \lambda\right).
\end{align}
Inserting equation~\eqref{eq:upq} and dropping factors of order unity we find
\begin{align}
	|\boldsymbol{u}\cdot\nabla\ln\rho| \approx \frac{u}{h}\left(\frac{h}{r} + \lambda\right),
\end{align}
hence
\begin{align}
	\left|\boldsymbol{\omega}_m\boldsymbol{u}\cdot\nabla\ln\rho\right| \approx \frac{u}{h}\left(|R\nabla\Omega| + \Omega\right)\left(\frac{h}{r} + \lambda\right).
\end{align}

Finally, we examine the term $\boldsymbol{u}\cdot\nabla\boldsymbol{\omega}_m$.
Again using $\boldsymbol{e}_p \approx \boldsymbol{e}_r$ and $\boldsymbol{e}_q \approx \boldsymbol{e}_\theta$ we find
\begin{align}
	\boldsymbol{u}\cdot\nabla\boldsymbol{\omega}_m
	&\approx \boldsymbol{e}_r \left[u_r \left(2\cos\theta\partial_r \Omega + \sin\theta \partial_r\partial_\theta\Omega\right)\right.\nonumber\\
	&\left. + \frac{u_\theta}{r}\left(3\cos\theta\partial_\theta\Omega+\sin\theta\left(\partial_\theta^2\Omega+r\partial_r\Omega\right)\right)\right]\nonumber\\
	&- \boldsymbol{e}_\theta \left[u_r \sin\theta\left(3\partial_r \Omega + r\partial_r^2\Omega\right)\right.\nonumber\\
	&\left. + \frac{u_\theta}{r}\left(r\cos\theta\partial_r\Omega+\sin\theta\left(\partial_\theta\Omega+r\partial_r\partial_\theta\Omega\right)\right)\right].
\end{align}
Making the same approximations as in equation~\eqref{eq:omu} we find
\begin{align}
	\left|\boldsymbol{u}\cdot\nabla\boldsymbol{\omega}_m\right| &\approx \frac{u}{h} \left[\Omega + |R\nabla\Omega|\right].
\end{align}

\subsection{Azimuthal Equation~\eqref{eq:vortp1}}

The first term is $R^{-1}\omega_\phi u_R$.
Expanding the azimuthal vorticity we find
\begin{align}
	\omega_\phi = \partial_r u_\theta + r^{-1} u_\theta - r^{-1} \partial_\theta u_r.
	\label{eq:vortexpPhi}
\end{align}
Noting that $u_\theta \approx u_q \approx u$ we may approximate the first term by $u/h$.
This is larger than the remaining terms, so
\begin{align}
	\omega_\phi \approx \frac{u}{h}.
	\label{eq:omegaPhiApprox}
\end{align}
Hence
\begin{align}
	R^{-1}\omega_\phi u_R
	&\approx \frac{u_R u}{h R}.
\end{align}
Averaged over latitudes $u_R$ projects comparably on to both $u_q$ and $u_p$, so $u_R \approx u$ and
\begin{align}
	|R^{-1}\omega_\phi u_R|
	&\approx \frac{u^2}{h R}.
\end{align}

The next term is $\omega_\phi\boldsymbol{u}\cdot\nabla\ln\rho$.
The first factor we have already evaluated in equation~\eqref{eq:omegaPhiApprox} while the latter we have computed in equation~\eqref{eq:lnr}, so
\begin{align}
	|\omega_\phi\boldsymbol{u}\cdot\nabla\ln\rho| \approx \frac{u^2}{h^2}|N|^2 \left(\frac{h}{r} + \lambda\right).
\end{align}
We have only incurred errors of order $h/r$ and $\lambda/\bar{\lambda}\approx \lambda$ in computing this term because we have approximated $u_\theta$ by $u$ so the expansion is accurate to leading order in both factors.

The next term is $\boldsymbol{u}\cdot\nabla\omega_\phi$.
Expanding the vorticity as in equation~\eqref{eq:vortexpPhi} we find
\begin{align}
	\boldsymbol{u}\cdot\nabla\omega_\phi
	&\approx \frac{\boldsymbol{e}_\phi}{r^2}\left[u_\theta\left(\partial_\theta u_\theta - \partial_\theta^2 u_r + r \partial_r \partial_\theta u_\theta\right)\right.\nonumber\\
	&\left.+u_r \left(\partial_\theta u_r - u_\theta + r\left(\partial_r u_\theta - \partial_r\partial_\theta u_r + r \partial_r^2 u_\theta\right)\right)\right].
\end{align}
This contains a term involving two radial derivatives of $u_\theta \approx u_q \approx u$ so that term dominates the expression abd we gave
\begin{align}
	|\boldsymbol{u}\cdot\nabla\omega_\phi| \approx \frac{u^2}{h^2}.
\end{align}

The next term is $\boldsymbol{\omega}\cdot\nabla(\Omega R)$.
Using equation~\eqref{eq:vortexp} we find
\begin{align}
	\boldsymbol{\omega}\cdot\nabla(\Omega R) &= \boldsymbol{\omega}_m \cdot \nabla(\Omega r \sin\theta) = \Omega \sin\theta \left(\partial_\theta\Omega\sin\theta - r\cos\theta \partial_r \Omega\right).
\end{align}
Neglecting the geometric factors this is just
\begin{align}
	|\boldsymbol{\omega}\cdot\nabla(\Omega R)| \approx \Omega|R\nabla\Omega|.
\end{align}

The final advective term is $\Omega \omega_R \boldsymbol{e}_\phi$.
Writing the vorticity in cylindrical coordinates we see that
\begin{align}
	\boldsymbol{\omega} = \boldsymbol{e}_R (-R \partial_z \Omega) + \boldsymbol{e}_z (2 \Omega + R \partial_R \Omega) + \boldsymbol{e}_\phi (\partial_z u_R - \partial_r u_z).
	\label{eq:vortC}
\end{align}
Hence,
\begin{align}
|\Omega \omega_R\boldsymbol{e}_\phi| &= R \Omega |\partial_z \Omega| \approx \Omega|R\nabla\Omega|.
\end{align}
%%%---------- close: stress_terms
%%%%%%%%%%% jump to inverse_cascade
%%%---------- open: inverse_cascade

\section{Inverse Cascade}
\label{sec:cascade}

In our analysis we have assumed that solutions to the governing equations contain no geometric factors which differ significantly from being of order unity.
In effect, we have assumed that any structures which form are in some fashion generic and do not depend specifically on $\Omega/|N|$, though they may depend on the regime in which the system lies.
While this assumption is usually sound, there is a known exception in the case of rapidly-rotating two-dimensional turbulence~\citep{1973BoLMe...4..345R,inverseArrest0}.
This phenomenon is known as the inverse cascade (also the Rhines or enstrophy cascade) and results from the Coriolis effect preferentially scattering waves into large-scale modes.
In both cases the result is feedback between small-scale convective motions and the overall geometry of the solution.

The reason that two-dimensional turbulence is relevant for our purposes is that stars and planets exhibit significant density stratification.
This makes turbulence effectively two-dimensional by restricting motion along the density gradient.

The Rhines cascade has been found both analytically and numerically~\citep{PhysRevE.65.067301} to result in the formation of alternating bands of differential rotation, also known as jets.
In particular, the number of jets is seen to scale as~\citep[see equation 21 of][]{2014Icar..237..143V}
\begin{align}
	n \approx \sqrt{\frac{\Omega R}{w}},
	\label{eq:numJets}
\end{align}
where we have used $d=R$ as the relevant vertical length-scale and $w$ is the characteristic velocity scale of turbulence in the system, given by $\varv_{\rm c}$ when the primary means of energy transport is convective.
A similar effect has been observed in MHD systems and leads to a similar scaling but with $w = \varv_{\rm A}$~\citep{diamond_itoh_itoh_silvers_2007}, though it leads to much weaker jets~\citep{2005PhPl...12e2515N}.
The appearance of jets and the scaling in equation~\eqref{eq:numJets} has been seen in a broad array of simulations~\citep{2013Icar..225..156G,2014Icar..237..143V} and agrees well with observations of the four gas giant planets in the solar system~\citep{1982Icar...52...62I,2001PhFl...13.1545G}.

When $n > 1$ the inverse cascade enhances latitudinal derivatives of quantities perturbed by the jets by a factor of $n$.
So for instance, if the stress $\mathbfss{T}$ is perturbed a fractional amount $\epsilon$ by the presence of jets then
\begin{align}
	\frac{1}{r}\frac{\partial \mathbfss{T}}{\partial \theta} \approx \frac{\mathbfss{T}}{r} + n \epsilon  \frac{\mathbfss{T}}{r}.
\end{align}
Similarly, it enhances radial derivatives such that
\begin{align}
	\frac{\partial \mathbfss{T}}{\partial r} \approx \frac{\mathbfss{T}}{h} + n \epsilon  \frac{\mathbfss{T}}{r}.
\end{align}
Taking $\epsilon$ to be no greater than $1$, we see that radial derivatives are only affected when $n > r / h \approx R / h$.
In the slowly-rotating limit our analysis was dominated by radial derivatives, generally by a factor of $R / h$, so at most this effect may serve to make the latitudinal derivatives comparably important.
We therefore expect no change to the scaling of any terms when $\Omega \la |N|$, because then $w \approx h |N|$ and $n < R/h$.

In the rapidly-rotating limit the situation is more complicated.
The dominant terms in the meridional vorticity and heat equations again contain radial derivatives, so any enhancement just rescales those equations and leaves their solution unchanged.

By contrast, in the azimuthal vorticity equation there are several terms which are not enhanced by the formation of jets.
These are the Taylor-Proudman term ($R \Omega\partial_z \Omega$) and the thermal wind term ($\nabla p \times \nabla \rho$).
The former is not enhanced because it involves only vertical ($\boldsymbol{e}_z$) derivatives, while the latter is not enhanced because in this limit it is already maximized with $\xi |N|^2 \approx |N|^2$.
The advective terms cannot be enhanced enough to dominate the azimuthal vorticity balance because they scale as $u^2 \la h^2 |N|^4/\Omega^2$, which decreases too rapidly with $\Omega$ to matter even after multiplying by an enhancing factor of $(n h / R)^2$.
Thus the only pieces which may be enhanced enough to matter are the stress terms.

We are not certain as to how the fluid stress responds to the introduction of a new, smaller length-scale.
It is possible that the scale of the stress is unchanged but its derivatives are enhanced, leading to a greater contribution and more differential rotation.
It is also possible that the stress is reduced in magnitude because its characteristic scale is shortened.
We favour the latter view because, far from the critical Reynolds or Rayleigh numbers, the only sensitive length-scales are those imposed by the geometry and background gradients of the system.

Taking this to be the case, we find that the stress terms with components in the cylindrical radial direction are diminished by the same factor of $n h / R$ by which derivatives are enhanced.
Therefore the only terms whose contributions to the vorticity equation are enhanced are those which contain more derivatives along $\boldsymbol{e}_R$ than stress indices along $\boldsymbol{e}_R$.
There are just two such terms which enter into the azimuthal vorticity equation as
\begin{align}
	&\partial_R^2 \mathbfss{T}_{Rz}\\
\intertext{and}
	&R^{-1} \partial_R \partial_\phi \mathbfss{T}_{\phi \phi}.
\end{align}
In each case there is one more radial derivative than radial index (subscript), so these contribute on the whole a factor of $n h / R$ more.

In the hydrodynamic limit $w \approx \varv_{\rm c} \approx h |N| k$ and the stress terms contribute of order $|N|^2 k^2$ to the azimuthal vorticity equation.
So after enhancement the stress contributes a net of
\begin{align}
	\frac{n h}{R}|N|^2 k^2 \approx \sqrt{\frac{h}{R}} |N|^{3/2} \Omega^{1/2} k^{3/2}.
\end{align}
The dominant terms are of order $|N|^2$, so the ratio of these enhanced terms to the dominant ones is
\begin{align}
	\frac{\sqrt{\frac{h}{R}} |N|^{3/2} \Omega^{1/2} k^{3/2}}{|N|^2} = \sqrt{\frac{h}{R}} |N|^{-1/2} \Omega^{1/2} k^{3/2}
\end{align}
Because $k$ falls at least as fast as $\Omega^{-1/2}$ and $h < R$ this ratio is less than unity, so there is no effect on our results in section~\ref{sec:fast_hydro}.

In the MHD limit the jets are driven by the Reynolds (non-magnetic) stress, not by the total turbulent stress~\citep[][see fig. 1.3 of]{diamond_itoh_itoh_silvers_2007}.
This means that
\begin{align}
\epsilon \approx \frac{\varv_{c}^2}{\varv_{\rm a}^2}.
\end{align}
Using equation~\eqref{eq:fastVC} we see that
\begin{align}
\epsilon  \approx \frac{k^2}{q^2}
\end{align}
The MHD stress contributes of order $q^2 |N|^2$ to the azimuthal vorticity equation, so the terms which are enhanced are of order $\epsilon q^2 |N|^2$.
The dominant terms are of order $|N|^2$, so the ratio of the enhanced terms to the dominant ones is $\epsilon q^2 \approx k^2$ times the enhancement factor.
With $w \approx \varv_{\rm A} \approx q h |N|$, this is
\begin{align}
	\frac{n h}{R} \approx \sqrt{\frac{h \Omega}{q R |N|}},
\end{align}
so the ratio of the enhanced terms to the dominant ones is of order
\begin{align}
k^2 \sqrt{\frac{h \Omega}{q R |N|}} \la \frac{h^{1/2} |N|^{3/2}}{R^{1/2}\Omega^{3/2}} \ll 1.
\end{align}
With $\Omega > |N|$ and $h < R$ we see that this is less than unity, so there is no effect on our results in section~\ref{sec:fast}.

%%%---------- close: inverse_cascade
%%%%%%%%%%% jump to breakup
%%%---------- open: breakup

\section{Breakup Rotation}
\label{sec:breakup}

Stable systems cannot rotate faster than the Keplerian velocity
\begin{align}
	\Omega_K \equiv \sqrt{\frac{g}{R}}
\end{align}
without invoking pressure profiles that increase outwards.
As a system approaches this velocity it nears the state of an accretion disk, in which
\begin{align}
	R\partial_z\Omega &= 0\\
	\intertext{and}
	R\partial_R\Omega &= -\frac{3}{2}\Omega.
\end{align}
The derivation of this state is straightforward, so we do not dwell on it.
However, we are interested in how this limit is approached.

To see why this is not simply an extension of the rapidly rotating limit, note that $|R\nabla\Omega|$ increases sub-linearly in that limit (equations~\ref{eq:hydroDR} and~\ref{eq:rapid_MHD_DR}), such that
\begin{align}
	|R\nabla\ln\Omega| \rightarrow 0
\end{align}
as $\Omega$ becomes large.
This is incompatible with the Keplerian limit of $-3/2$ and so something different must happen in between the two limits.

The first important point is that the window of rotation rates we considered in section~\ref{sec:fast} is not infinite.
In particular, we considered 
\begin{align}
	|N| \ll \Omega \ll \sqrt{\frac{g}{R}} = \Omega_K,
\end{align}
so the window has width
\begin{align}
	\frac{\Omega_\mathrm{max}}{\Omega_\mathrm{min}} = \sqrt{\frac{g}{R|N|^2}}.
\end{align}
For $\Omega > \Omega_{\rm max}$ certain terms which could be ignored because $h \ll r$ become significant however this does not explain the difference between our results and the Keplerian limit because such terms do not scale sufficiently quickly with $\Omega$ as to produce $|R\nabla\Omega| \propto \Omega$.

The second point to note is that, as the rotation rate increases, so does $|N|$.
To incorporate both the hydrodynamic and the magnetized limits discussed in the previous section we write
\begin{align}
	|N| = |N|_0 \left(\frac{\Omega}{|N|_0}\right)^{\alpha}
\end{align}
for some $\alpha > 0$.
Using equation~\eqref{eq:brvs} we may relate this to the entropy gradient and find that
\begin{align}
	|\nabla s| = \frac{|N|_0^2}{g \bar{\xi}} \left(\frac{\Omega}{|N|_0}\right)^{2\alpha},
	\label{eq:ns}
\end{align}
where $\boldsymbol{g}$ is the effective gravitational field accounting for the centrifugal acceleration and the factor of $\bar{\xi}$ captures the misalignment between this and the entropy gradient.
In the simple case of an ideal gas with aligned pressure and density gradients this may be related to the temperature gradient using equation~\eqref{eq:s}, so
\begin{align}
	|\nabla s| &= |\nabla \ln p - \gamma \nabla \ln \rho|\\
	&= |(1-\gamma)\nabla \ln p + \gamma \nabla \ln T|\\
	&= \frac{1}{h} \left[1 - \gamma + \gamma \frac{d\ln T}{d\ln p}\right]\\
	&= \frac{\gamma}{h} \left[\nabla - \nabla_{\mathrm{a}}\right],
	\label{eq:grad_s_breakup}
\end{align}
where $\nabla_{\mathrm{a}}$ and $\nabla$ are the adiabatic and actual logarithmic temperature gradients respectively.
But $\nabla$ is bounded above by the radiative temperature gradient $\nabla_{\mathrm{R}}$, so
\begin{align}
	|\nabla s| \leq \frac{\gamma}{h} \left[\nabla_{\mathrm{R}} - \nabla_{\mathrm{a}}\right].
	\label{eq:ns2}
\end{align}
Combining equations~\eqref{eq:ns} and~\eqref{eq:ns2} we find
\begin{align}
	|N|_0^2 \left(\frac{\Omega}{|N|_0}\right)^{2\alpha} \leq \frac{\gamma g \bar{\xi}}{h} \left[\nabla_{\mathrm{R}} - \nabla_{\mathrm{a}}\right].
\end{align}
Using equation~\eqref{eq:h} and recalling that for an ideal gas $c_\mathrm{s}^2 = \gamma p/\rho$ we see that
\begin{align}
	|N|_0^2 \left(\frac{\Omega}{|N|_0}\right)^{2\alpha} \leq \bar{\xi}\left(\frac{\gamma g}{c_\mathrm{s}}\right)^2 \left[\nabla_{\mathrm{R}} - \nabla_{\mathrm{a}}\right].
	\label{eq:ineq0}
\end{align}
To proceed note that $g$ in this equation is the effective\footnote{i.e. centrifugally-corrected} gravity, such that
\begin{align}
	\boldsymbol{g} = \boldsymbol{g}_0 - \boldsymbol{e}_R \Omega^2 R
	\label{eq:gmod}
\end{align}
where $\boldsymbol{g}_0$ is the actual acceleration of the gravitational interaction.
Hence, on the equator, where $\boldsymbol{g}_0$ is parallel to $\boldsymbol{e}_R$,
\begin{align}
	g = g_0 \left(1 - \frac{\Omega^2 R}{g_0}\right).
\end{align}
So
\begin{align}
	\bar{\xi}\left(\frac{\gamma g_0}{c_\mathrm{s} |N|_0}\right)^2 \left(1 - \frac{\Omega^2 R}{g_0}\right)^2\left(\frac{\Omega}{|N|_0}\right)^{-2\alpha}  \left[\nabla_{\mathrm{R}} - \nabla_{\mathrm{a}}\right] \geq 1.
	\label{eq:ineq1}
\end{align}
As $\Omega$ increases this breaks down because the left-hand side vanishes, both because $\Omega^{-2\alpha}$ and as a result of the centrifugal term.
In other words at some point the system must become radiative.
Of course this is only a problem in regions near the equator but it may have significant consequences for the transport of angular momentum and hence may suffice to explain the discrepancy between the Keplerian and sub-Keplerian regimes.

%%%---------- close: breakup

\bsp	% typesetting comment
\label{lastpage}
\end{document}